\documentclass[twocolumn]{aastex63}

\graphicspath{{./}{Figures/}}

\received{December 27, 2019}
\accepted{February 14, 2020}
\shorttitle{HSC-SSP Early-phase SNe Ia}
\shortauthors{J. Jiang et al.}
\begin{document}

\title{The HSC-SSP Transient Survey: Implications from Early Photometry and Rise Time of Normal Type Ia Supernovae}

\correspondingauthor{Ji-an Jiang}
\email{jian.jiang@ipmu.jp}

\author[0000-0002-9092-0593]{Ji-an Jiang}
\affil{Kavli Institute for the Physics and Mathematics of the Universe (WPI), The University of Tokyo Institutes for Advanced Study, The University of Tokyo, 5-1-5 Kashiwanoha, Kashiwa, Chiba 277-8583, Japan}

\author{Naoki Yasuda}
\affiliation{Kavli Institute for the Physics and Mathematics of the Universe (WPI), The University of Tokyo Institutes for Advanced Study, The University of Tokyo, 5-1-5 Kashiwanoha, Kashiwa, Chiba 277-8583, Japan}

\author[0000-0003-2611-7269]{Keiichi Maeda}
\affiliation{Department of Astronomy, Kyoto University, Kitashirakawa-Oiwake-cho, Sakyo-ku, Kyoto 606-8502, Japan}
\affiliation{Kavli Institute for the Physics and Mathematics of the Universe (WPI), The University of Tokyo Institutes for Advanced Study, The University of Tokyo, 5-1-5 Kashiwanoha, Kashiwa, Chiba 277-8583, Japan}

\author{Mamoru Doi}
\affiliation{Institute of Astronomy, Graduate School of Science, The University of Tokyo, 2-21-1 Osawa, Mitaka, Tokyo 181-0015, Japan}
\affiliation{Research Center for the Early Universe, Graduate School of Science, The University of Tokyo, 7-3-1 Hongo, Bunkyo-ku, Tokyo 113-0033, Japan}
\affiliation{Kavli Institute for the Physics and Mathematics of the Universe (WPI), The University of Tokyo Institutes for Advanced Study, The University of Tokyo, 5-1-5 Kashiwanoha, Kashiwa, Chiba 277-8583, Japan}

\author[0000-0002-4060-5931]{Toshikazu Shigeyama}
\affiliation{Research Center for the Early Universe, Graduate School of Science, The University of Tokyo, 7-3-1 Hongo, Bunkyo-ku, Tokyo 113-0033, Japan}

\author[0000-0001-8537-3153]{Nozomu Tominaga}
\affiliation{Department of Physics, Faculty of Science and Engineering, Konan University, 8-9-1 Okamoto, Kobe, Hyogo 658-8501, Japan}
\affiliation{Kavli Institute for the Physics and Mathematics of the Universe (WPI), The University of Tokyo Institutes for Advanced Study, The University of Tokyo, 5-1-5 Kashiwanoha, Kashiwa, Chiba 277-8583, Japan}

\author[0000-0001-8253-6850]{Masaomi Tanaka}
\affiliation{Astronomical Institute, Tohoku University, Aoba, Sendai 980-8578, Japan}
\affiliation{Kavli Institute for the Physics and Mathematics of the Universe (WPI), The University of Tokyo Institutes for Advanced Study, The University of Tokyo, 5-1-5 Kashiwanoha, Kashiwa, Chiba 277-8583, Japan}

\author[0000-0003-1169-1954]{Takashi J. Moriya}
\affiliation{National Astronomical Observatory of Japan, National Institutes of Natural Sciences, 2-21-1 Osawa, Mitaka, Tokyo 181-8588, Japan}
\affiliation{School of Physics and Astronomy, Faculty of Science, Monash University, Clayton, VIC 3800, Australia}

\author{Ichiro Takahashi}
\affiliation{Kavli Institute for the Physics and Mathematics of the Universe (WPI), The University of Tokyo Institutes for Advanced Study, The University of Tokyo, 5-1-5 Kashiwanoha, Kashiwa, Chiba 277-8583, Japan}
\affiliation{CREST, JST, 4-1-8 Honcho, Kawaguchi, Saitama 332-0012, Japan}

\author[0000-0001-7266-930X]{Nao Suzuki}
\affiliation{Kavli Institute for the Physics and Mathematics of the Universe (WPI), The University of Tokyo Institutes for Advanced Study, The University of Tokyo, 5-1-5 Kashiwanoha, Kashiwa, Chiba 277-8583, Japan}

\author{Tomoki Morokuma}
\affiliation{Institute of Astronomy, Graduate School of Science, The University of Tokyo, 2-21-1 Osawa, Mitaka, Tokyo 181-0015, Japan}
\affiliation{Kavli Institute for the Physics and Mathematics of the Universe (WPI), The University of Tokyo Institutes for Advanced Study, The University of Tokyo, 5-1-5 Kashiwanoha, Kashiwa, Chiba 277-8583, Japan}

\author[0000-0001-9553-0685]{Ken'ichi Nomoto}
\affiliation{Kavli Institute for the Physics and Mathematics of the Universe (WPI), The University of Tokyo Institutes for Advanced Study, The University of Tokyo, 5-1-5 Kashiwanoha, Kashiwa, Chiba 277-8583, Japan}




\vspace{10pt}

\begin{abstract}

With a booming number of Type Ia supernovae (SNe Ia) discovered within a few days of their explosions, a fraction of SNe Ia that show luminosity excess in the early phase (early-excess SNe Ia) have been confirmed. In this article, we report early-phase observations of seven photometrically normal SNe Ia (six early detections and one deep non-detection limit) at the COSMOS field through a half-year transient survey as a part of the Hyper Suprime-Cam Subaru Strategic Program (HSC SSP). In particular, a blue light-curve excess was discovered for HSC17bmhk, a normal SN Ia with rise time longer than 18.8 days, during the first four days after the discovery. The blue early excess in optical wavelength can be explained not only by interactions with a non-degenerate companion or surrounding dense circumstellar matter but also radiation powered by radioactive decays of $^{56}$Ni at the surface of the SN ejecta. Given the growing evidence of the early-excess discoveries in normal SNe Ia that have longer rise times than the average and a similarity in the nature of the blue excess to a luminous SN Ia subclass, we infer that early excess discovered in HSC17bmhk and other normal SNe Ia are most likely attributed to radioactive $^{56}$Ni decay at the surface of the SN ejecta. In order to successfully identify normal SNe Ia with early excess similar to that of HSC17bmhk, early UV photometries or high-cadence blue-band surveys are necessary.

\end{abstract}



\section{Introduction} \label{sec:intro}

Type Ia supernovae (SNe Ia) are thought to originate from the thermonuclear explosion of a carbon-oxygen white dwarf (WD) in a binary system. The specific photometric behavior of SNe Ia makes them an important cosmic distance indicator \citep{perlmutter97,perlmutter99,riess98}, while the progenitor and the mechanism leading to the explosion are still under debate. In the last decade, with a booming number of SNe Ia discovered within a few days of their explosions, remarkable work on such long-standing issues has been done by investigating very early photometric behavior of some SNe Ia \citep{li11,nugent11,cao15,JJA2017}.

Theoretically, \citet{kasen10} proposed that a prominent brightening in the first few days of the explosion can be observed under specific viewing directions due to the interaction between the expanding ejecta and a non-degenerate companion star, which makes SNe Ia with additional luminosity enhancement in the early time a powerful indicator of the single-degenerate progenitor scenario \citep{kasen10,maeda14,kutsuna15}. In addition to the companion-interaction scenario, an interaction between dense circumstellar matter (CSM) and SN ejecta (``CSM-ejecta interaction"; \citealp{levanon15}, \citeyear{levanon17}) and vigorous mixing of radioactive $^{56}$Ni in the outermost region of SN ejecta (``surface-$^{56}$Ni-decay"; \citealp{piro16}) may produce similar early light-curve excess to that predicted by the companion interaction. Moreover, radiation from short-lived radioactive elements generated by a precursory detonation at a helium shell of the primary WD (``He-shell detonation"; He-det) can cause a prominent but relatively red early excess \citep{JJA2017,maeda18}. In this article, we define ``early-excess SNe Ia (EExSNe Ia)" as SNe Ia that show luminosity excess compared to ``smoothly rising" SNe Ia whose light curves can be roughly fitted by single power-law functions, such as the classical fireball model with a power-law index of 2, in the first few days of explosions (see Figure 2 of \citet{JJA2018}).

Several EExSNe Ia have been discovered in recent years \citep{cao15,smitka15,marion16,JJA2017,hosseinzadeh17,miller18}. Interestingly, as opposed to early-excess discoveries in 91T/99aa-like luminous SNe Ia, which show extremely high early-excess fraction likely due to an efficient detonation reaching to the outermost region of their progenitors under the surface-$^{56}$Ni-decay scenario (\citealp{JJA2018}), only three early-excess events were reported for normal SNe Ia (SN~2017cbv, \citealp{hosseinzadeh17}; SN~2017erp, \citealp{JJA2018,brown19}; SN~2018oh, \citealp{shappee19,dimitriadis19,li19})\footnote{Note that SN~2017cbv and SN~2018oh show relatively shallow Si \textsc{ii} absorptions in the early phase and luminous light curves compared with those of typical normal SNe Ia such as SN~2011fe. However, given the similar spectral evolution from one week before the peak among the three objects as well as larger ${\Delta}m_{15}$ values than those of 91T/99aa-like luminous SNe Ia, we here classify SN~2017cbv and SN~2018oh as ``normal" SNe Ia, following the argument of previous papers.}, and origins of the excess are under debate. The rare detections of early excess in normal SNe Ia can be attributed to intrinsically inconspicuous early-excess features or a small observable viewing-angle range under specific early-excess scenarios.

The rise time (i.e. the time from first light to $B$-band maximum brightness) of SNe Ia can be well estimated by the early-phase observation. Statistically, stretch-corrected mean rise time was found to be 17--18 days based on large samples of SNe Ia \citep{hayden10,bianco11,ganeshalingam11,gonzalez12}, which is in line with studies of individual normal SNe Ia discovered at very early time \citep{foley11,nugent11,zheng13}. With an even larger sample of early SNe Ia, a large scatter of rise times with a mean value of 18.5 days (without stretch correction) was reported by the Zwicky Transient Factory (ZTF) very recently \citep{miller20}. In contrast, EExSNe Ia usually show longer rise times than the mean values obtained from statistics \citep{cao15,marion16,JJA2017}, suggesting the existence of a long dark phase of some SNe Ia \citep{piro13,mazzali14}.

In this paper, we report seven photometrically normal SNe Ia with multiband early observations or deep non-detection constraints discovered by a transient survey as a part of the Hyper Suprime-Cam Subaru Strategic Program (``the HSC-SSP transient survey"). One of the seven SNe shows a clear early-excess feature. The paper is organized as follows. A brief review of the HSC-SSP transient survey and sample selection are presented in \S2. Photometric behavior and basic analyses of HSC-SSP early-phase SNe Ia are summarized in \S3. Further discussions and conclusions are given in \S4.

\section{The HSC-SSP Transient Survey}

\subsection{Observation, data reduction, and transient identification}

Observations for this study were carried out as a part of the Hyper Suprime-Cam Subaru Strategic Program (HSC SSP; \citealp{miyazaki18,aihara18}) from November 2016 to April 2017 at the COSMOS field (HSC-SSP UltraDeep layer) and adjacent areas (HSC-SSP Deep layer). The basic observing strategy is to take images at two epochs separated by 7--10 days in each of five broad bands of HSC ($g$, $r$, $i$, $z$, and $y$ bands; \citet{kawanomoto18}) during monthly observing runs for the UltraDeep layer. In total, we carried out photometries at 12 epochs over six months in five bands. For the Deep layer, observations were carried out six times over four months, with shorter exposure time than at the UltraDeep observations. An observing log was summarized in Table 1 of \citet{yasuda19}.

Standard image reduction procedures including bias, dark, flat, and fringe corrections, as well as astrometric and photometric calibrations against the Pan-STARRS1 (PS1) $3\pi$ catalog \citep{tonry12,magnier13} were done with the HSC pipeline, a version of the Large Synoptic Survey Telescope (LSST) stack \citep{axelrod10,bosch18,ivezic19}. The image subtraction was applied using deep, coadded reference images created from data taken in March 2014, April 2016, and May 2019\footnote{For the filter uniformity between observational data and reference, the $i$-band photometry was performed on subtracted images using new $i$-band references taken on May 10th, 2019.}. The forced point-spread function photometry was performed at the location of objects left in the reference-subtracted images. We refer to \citet{aihara18} and \citet{yasuda19} for further details on the HSC data reduction.

SN candidates were identified from 65,387 transients selected by the machine-learning technique. We firstly filtered out objects with ``negative" detections (e.g. faded transients) or with point-source hosts at the same location (e.g. variable stars). Then, each of the remaining 26,988 candidates were cross-checked by team members. With specific criteria (e.g. the location relative to the host, X-ray association, light-curve shape, etc.), finally, 1824 objects were classified as SNe and 1534 as active galactic nuclei (AGNs). Marginal cases (139 objects) were tagged as ``SN/AGN" \citep{yasuda19}.

\subsection{SN Ia selection}

The SALT2 light-curve fitting code \citep{guy10,betoule14} was applied to our SN sample for the SN Ia classification. All SNe were fitted with snfit (ver 2.4) by using $g$-, $r$-, $i$-, and $z$-band photometric data\footnote{Since $y$-band images suffer from scattered light, and some of the objects were affected by imperfect correction \citep{aihara18}, the light-curve fitting did not use $y$-band data to avoid systematical error from an inhomogeneous classification process.} and fixing the redshift parameter upon the most reliable redshift information, i.e. either spectroscopic redshifts (spec-$z$) or COSMOS2015 photometric redshifts (COSMOS photo-$z$, \citealp{laigle16}). Different classification methods were used for those which do not associate with a clear host galaxy (i.e. hostless SNe) or only have photometric redshifts from HSC broadband photometry \citep{yasuda19}. Further details about the SN classification is prepared in a separated paper (Takahashi, I., et al. 2020, in preparation).

Given that the main scientific goal of the HSC-SSP transient survey is high-$z$ SN Ia cosmology, the following criteria were applied to select HSC-SSP SNe Ia samples: (1) light-curve parameters, shape ($x_1$) and color ($c$) within the 3$\sigma$ range of \citet{scolnic16}'s ``All G10" distribution; (2) a fitting $B$-band peak absolute magnitude, $M_B$ $\lesssim$ $-18.5$ mag ($H_{0}=70.4$ km s$^{-1}$ Mpc$^{-1}$, $\Omega_M=0.272$); (3) a reduced $\chi^2$ $<$ 10 with the number of degrees of freedom (ndf) $\geq$ 5. In total, 433 SNe were classified as SNe Ia, and 129 of them have host spec-$z$ or COSMOS photo-$z$ information. Note that the relatively small fraction of SNe Ia compared with total 1824 SN candidates is due to our strict criteria not only to rule out peculiar SN Ia subclasses (e.g. subluminous SNe Ia) but also to omit SNe Ia with imperfect fittings and/or limited photometric data points. Finally, six SNe Ia with detection magnitudes and one SN Ia with early non-detection limiting magnitude $> 3.2$ mag fainter than their peaks in single/multiple bands were labeled as HSC-SSP early-phase SNe Ia\footnote{There are an additional three SNe Ia with early observations or deep early non-detections in our sample. The photometric information is, however, limited for these SNe to pin down their rise times; we thus focus on the best-observed seven objects through the paper.}. We only used observed light curves (with time-dilation correction) and compared them with numerical simulations to avoid additional uncertainties by applying K-correction on the early-phase data.

\section{HSC-SSP Early-phase SNe Ia}

Even though a low-cadence survey gazing at a few square degrees of the sky is not optimized for searching early-phase SNe Ia, thanks to the deep imaging capability of Subaru/HSC, six photometrically normal early-phase SNe Ia, including an EExSN Ia, HSC17bmhk, were discovered at the COSMOS field and adjacent areas through the half-year HSC-SSP transient survey. In addition to the detections of the six normal SNe Ia in very early time, a deep non-detection is available for another normal SN Ia, HSC17dbjm, which also can be used to constrain its early light-curve behavior. Basic information of the seven HSC-SSP early-phase SNe Ia is listed in Table 1. Redshifts of the seven objects are between 0.25 and 0.48, which were derived from either spectra (6/7) or COSMOS photo-$z$ (1/7) of their host galaxies. Given the high redshifts of the host galaxies, distances of the HSC-SSP early-phase SNe Ia were determined without taking into account hosts' peculiar velocities. In addition, the high redshifts of the HSC-SSP early-phase SNe Ia enabled us to study their near-ultraviolet light curves from early time. Nevertheless, as the HSC-SSP transient survey is designed for high-$z$ SN cosmology, most of the spectroscopic time was used for the host galaxies to confirm SN redshifts after the survey rather than for SNe themselves. Thus, we did not get spectral information of the HSC-SSP early-phase SNe Ia.

\begin{figure*}
\gridline{\fig{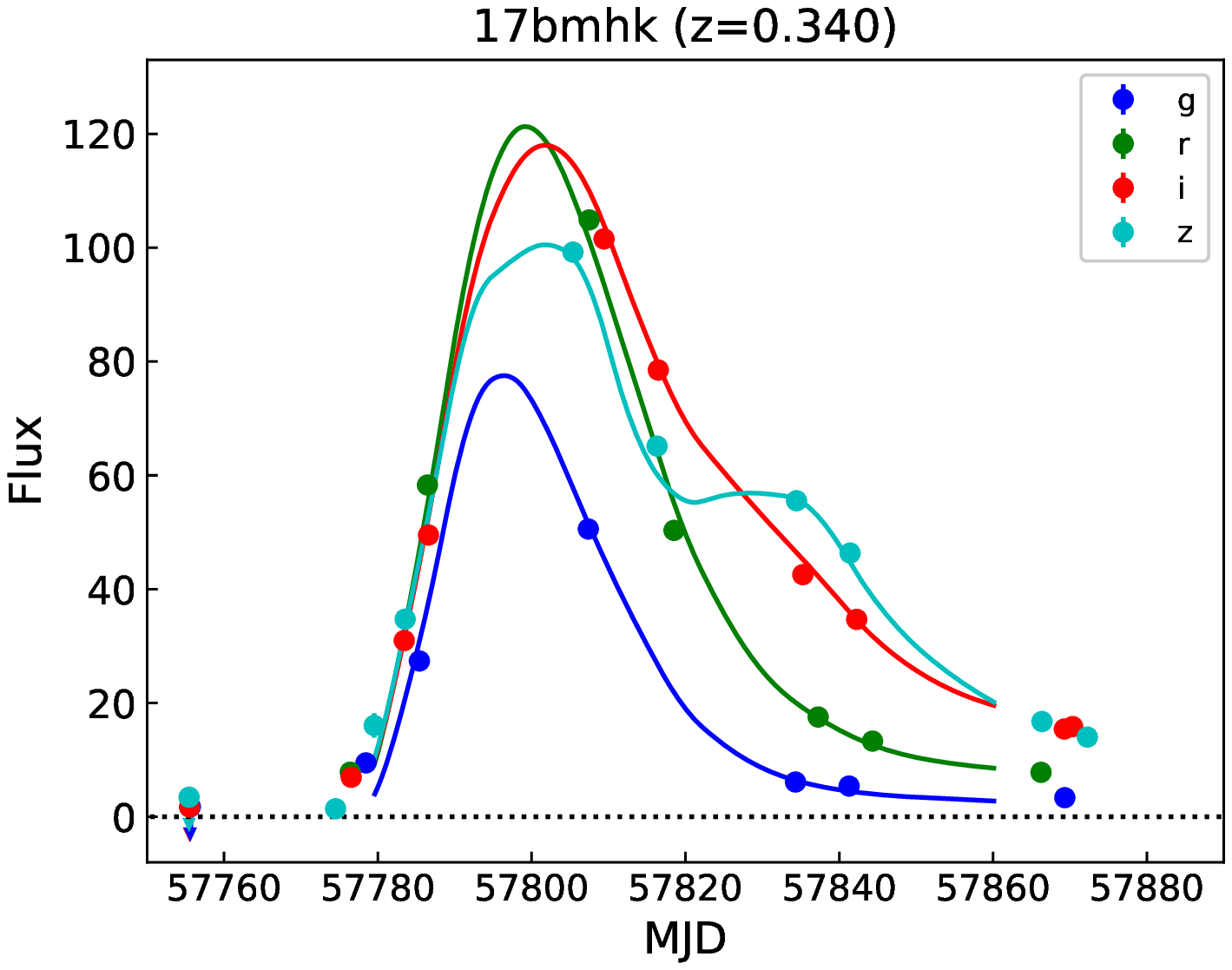}{0.6\textwidth}{(a)}
          }
\gridline{\fig{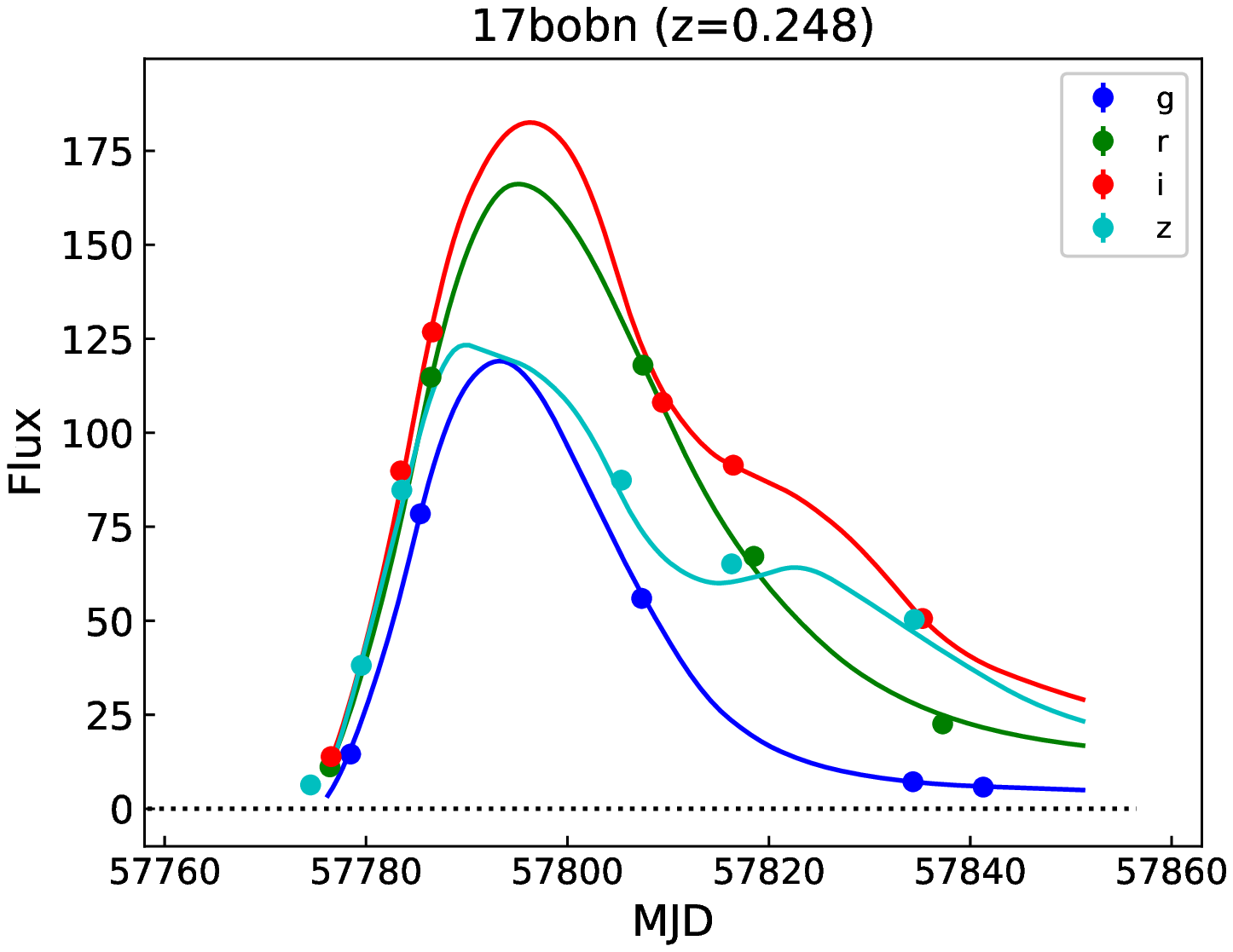}{0.34\textwidth}{(b)}
          \fig{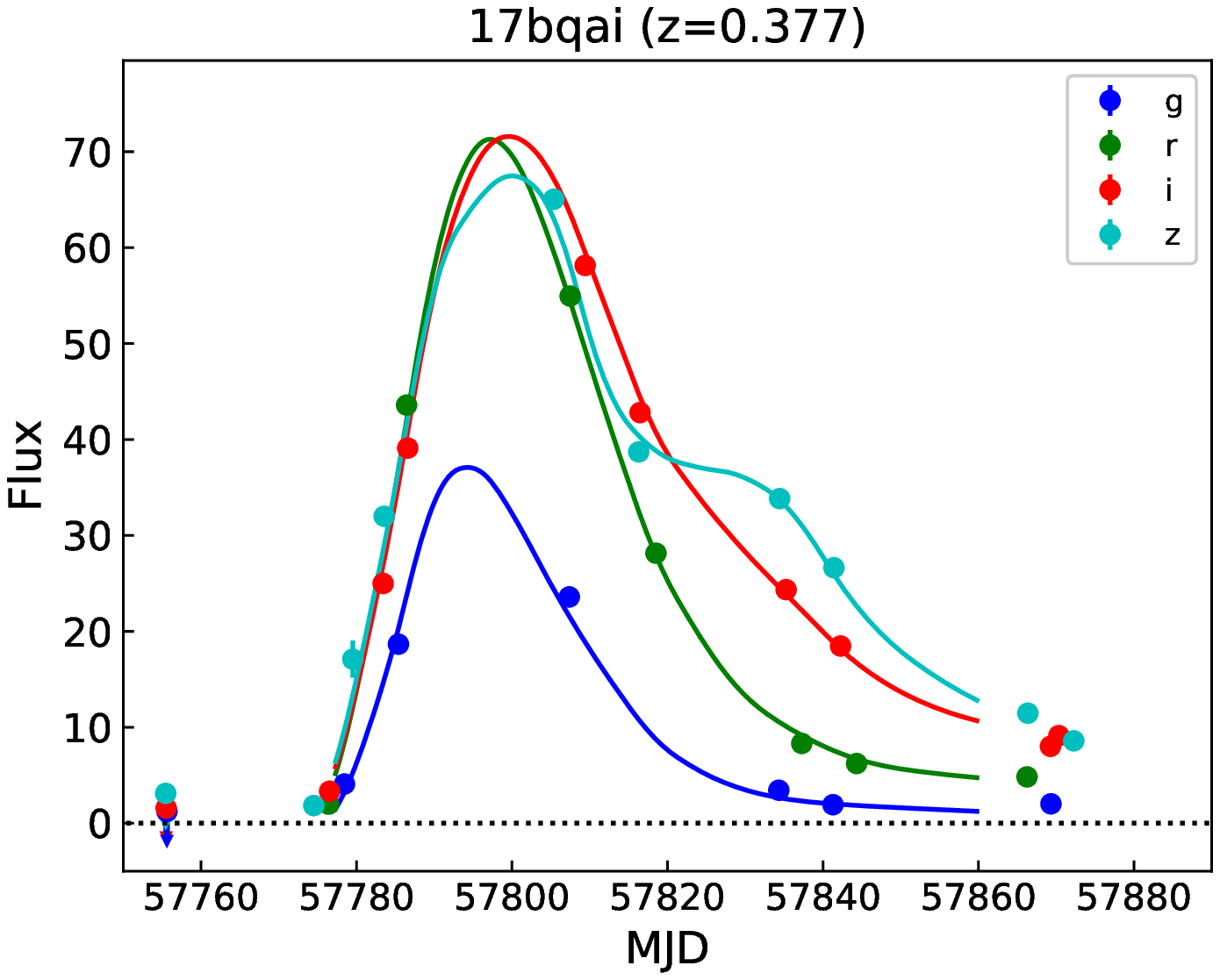}{0.34\textwidth}{(c)}
          \fig{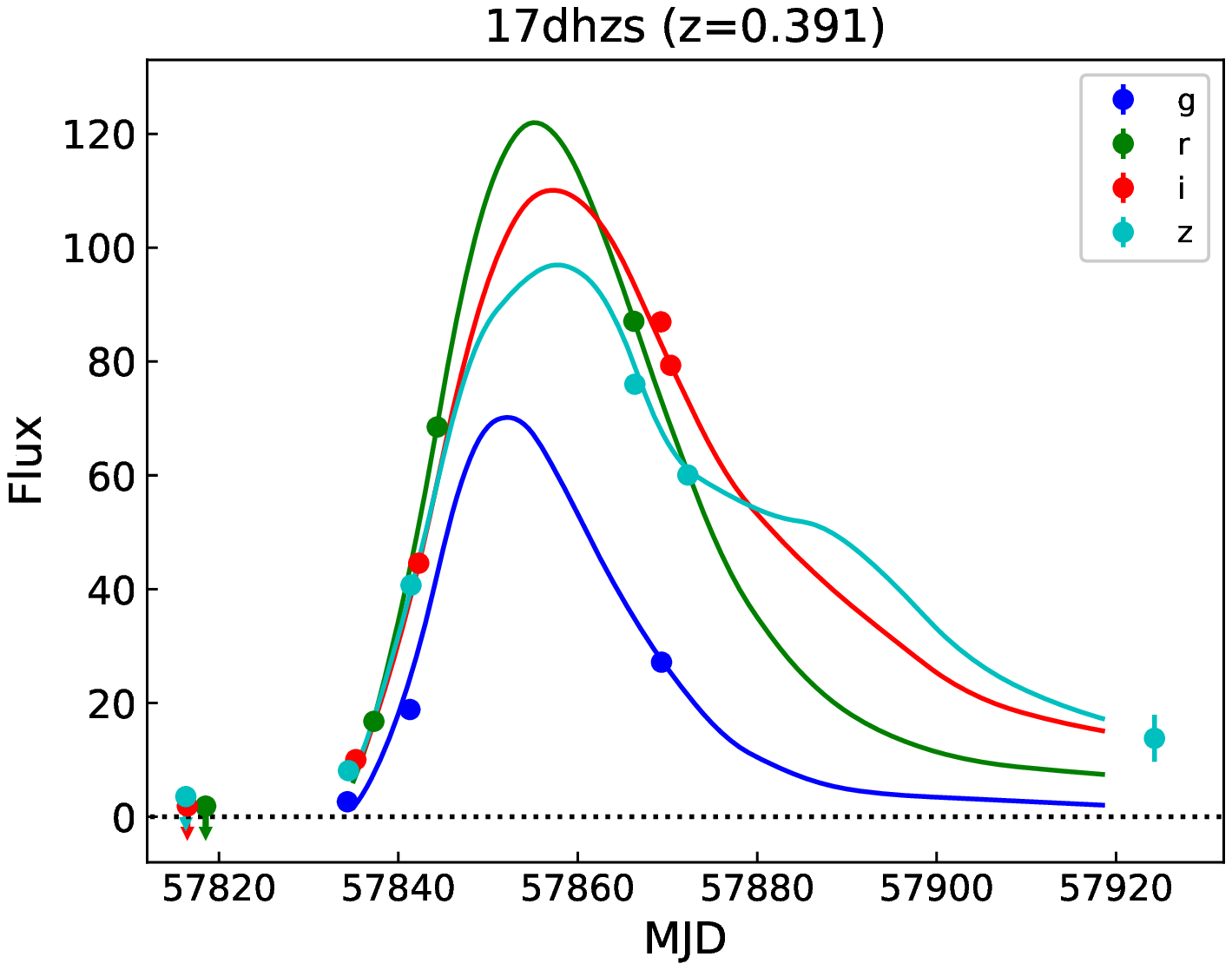}{0.34\textwidth}{(d)}
          }
\gridline{\fig{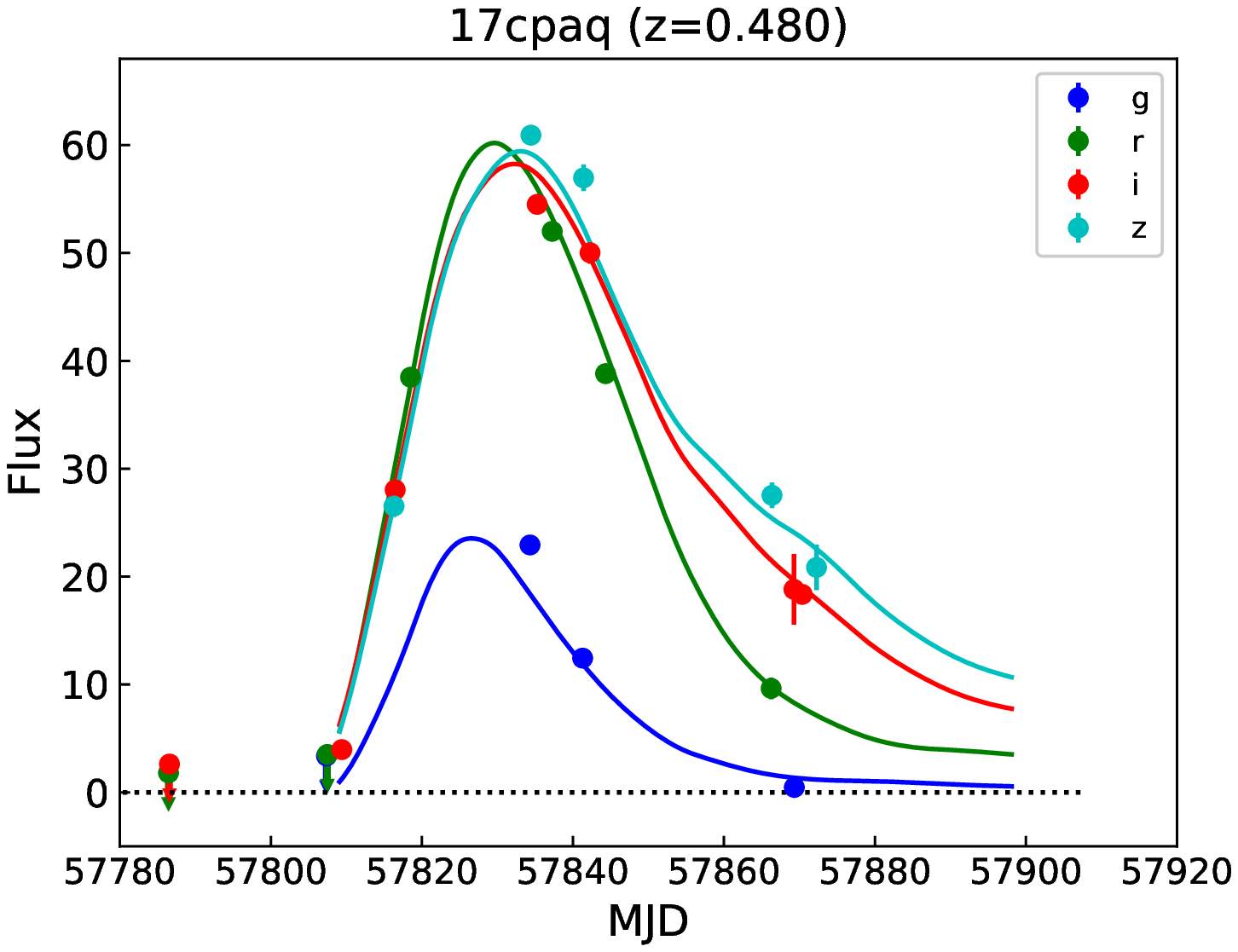}{0.34\textwidth}{(e)} 
          \fig{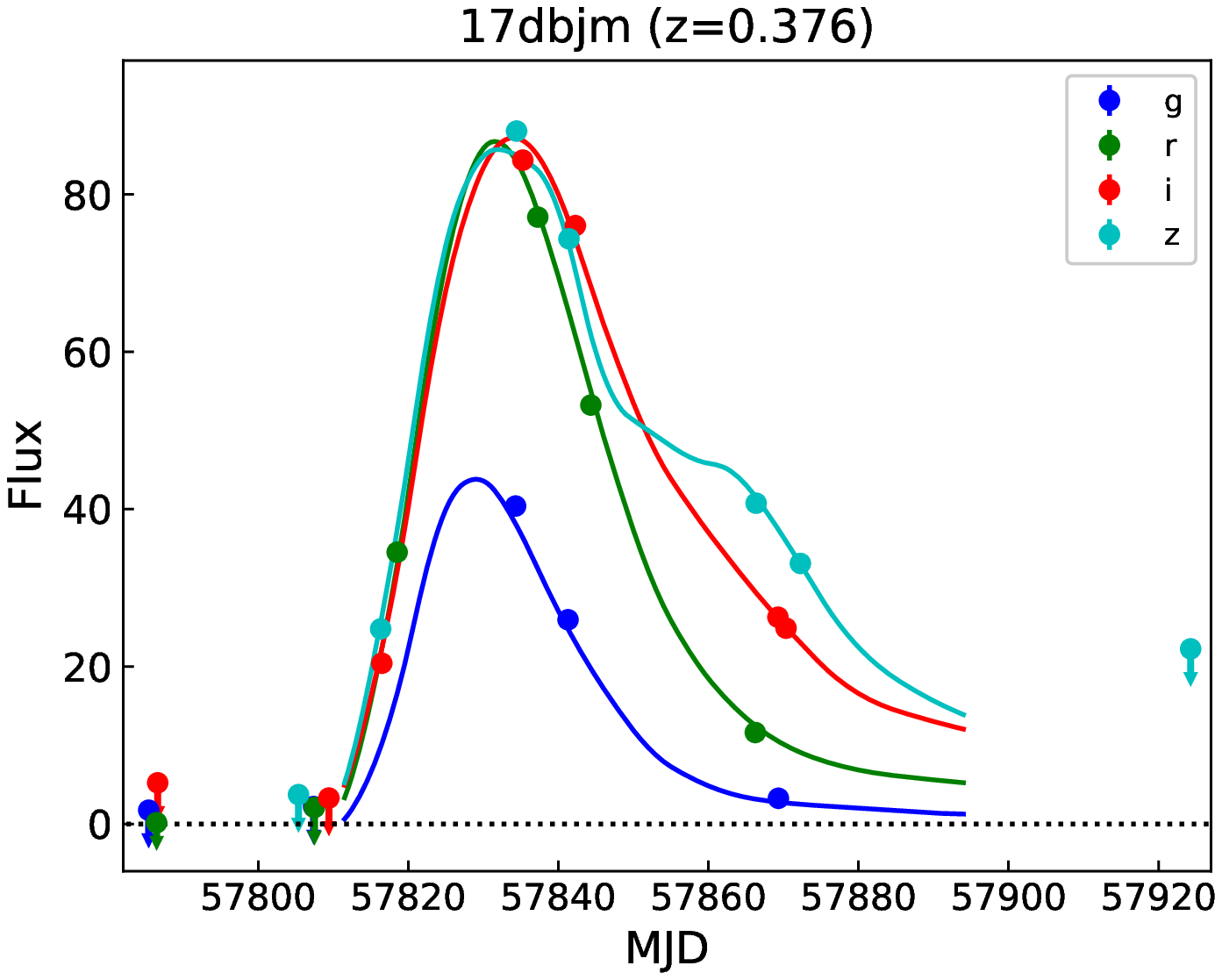}{0.34\textwidth}{(f)}
          \fig{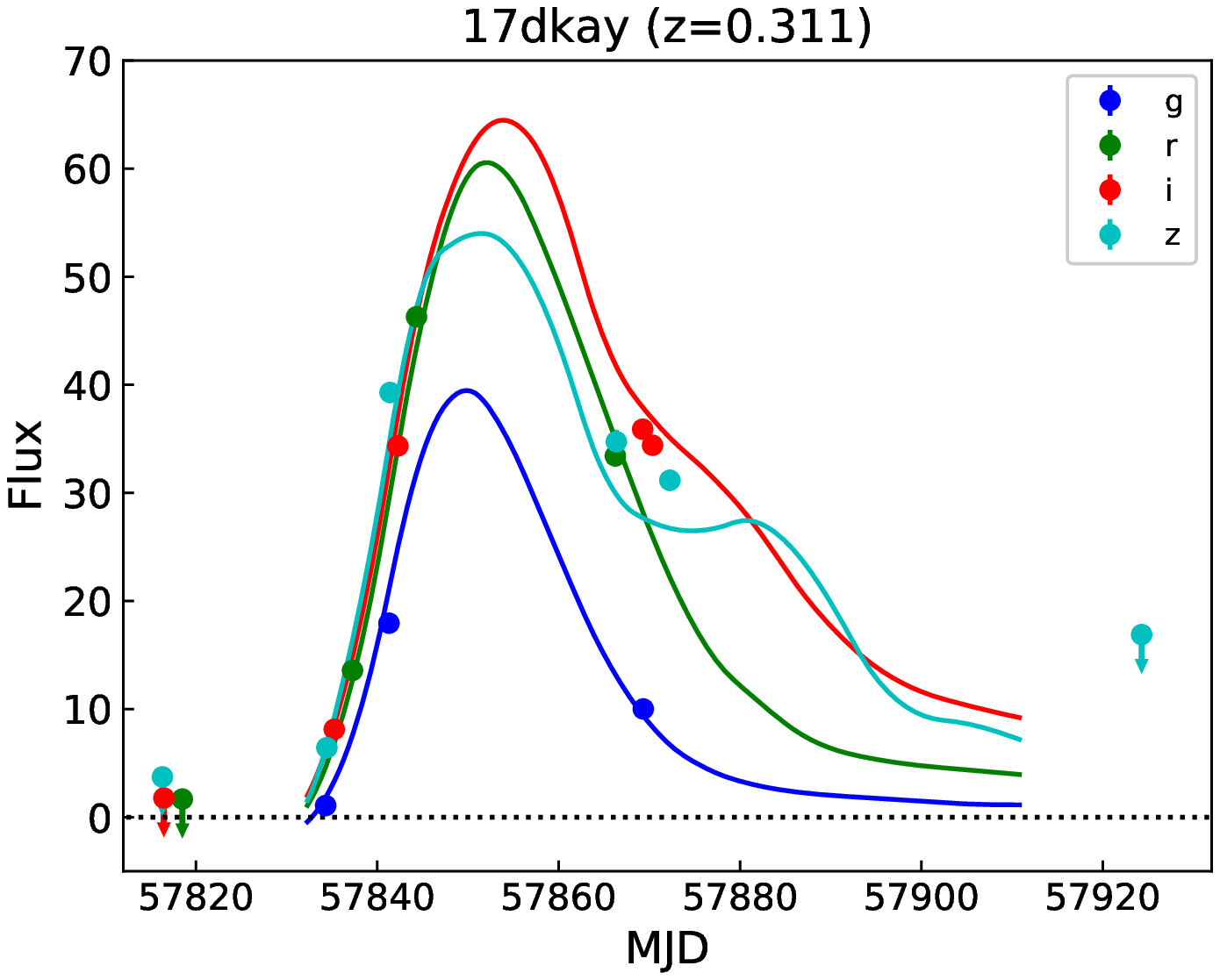}{0.34\textwidth}{(g)}
          }
\caption{Observed light curves and the best-fit SALT2 models (solid curves) of HSC-SSP early-phase SNe Ia. Error bars show 1$\sigma$, and points with arrows are 5$\sigma$ non-detections. The zero-point magnitude is set to 27.0 mag. Fitting parameters of each object are given in Table 2. \label{fig:LightCurves}}
\end{figure*}

\begin{deluxetable*}{ccccccccc}
\centerwidetable
\tablenum{1}
\tablecaption{Characteristics of HSC-SSP Early-phase SNe Ia\label{tab:HSC-SSP Early-phase SNe Ia}}
\tablewidth{0pt}
\tablehead{
\colhead{Name} & \colhead{$M_B$ $^a$} & \colhead{${\Delta}m_{15}(B)$} & \colhead{Redshift $^b$} & \colhead{Rise Time (days) $^c$} & \colhead{$r-i$ $^d$} & \colhead{$i-z$ $^d$} & \colhead{$g-z$ $^d$}
}
\startdata
HSC17bmhk & $-19.14$(03) & 1.04(01) & 0.340 & $>18.77 (18.64)$ & $-0.14(10)$ @$t=-17.3$ & 0.11(04) @$t=-12.1$ & -- \\
\hline
HSC17bqai & $-18.84$(03) & 1.06(01) & 0.377 & $>17.04 (17.07)$ &  $0.55(29)$ @$t=-15.6$ & 0.26(04) @$t=-10.5$ & -- \\
HSC17dbjm & $-19.17$(03) & 1.18(01) & 0.376 & $\sim16.49 (17.72)$ & -- & $0.20(06)$ @$t=-11.4$ & -- \\
HSC17bobn & $-18.99$(03) & 1.15(02) & 0.248 & $\sim17.07 (17.98)$ & $0.22(11)$ @$t=-14.7$ & -- & -- \\
HSC17cpaq & $-19.15$(03) & 0.97(02) & 0.480 & $\sim16.86 (16.00)$ & -- & $-0.07(07)$ @$t=-10.1$ & -- \\
HSC17dhzs & $-19.27$(05) & 1.13(05) & 0.391 & $\sim17.15 (18.12)$ & -- & -- & $1.18(26)$ @$t=-15.5$ \\
HSC17dkay & $-18.53$(04) & 1.31(04) & 0.311 & $\sim15.84 (18.15)$ & -- & -- & $1.89(39)$ @$t=-13.4$ \\
\enddata
\tablecomments{\\
$^a$ The peak absolute magnitude in the rest-frame $B$ band, which is derived from the best-fit SALT2 model. Numbers in parentheses are 1$\sigma$ uncertainties in units of 0.01 magnitude.\\
$^b$ Redshift information derived from specific emission lines of host galaxies except for HSC17dkay, whose redshift is adopted from the COSMOS photo-$z$ \citep{laigle16}.\\
$^c$ Rise times of HSC17bmhk and HSC17bqai are defined as rest-frame times from HSC discoveries to their $B$-band peaks, which are shorter than the real value. Rise times of other HSC-SSP early-phase SNe Ia are obtained from the extrapolated power-law fittings. Numbers in parentheses are stretch-corrected rise times.\\
$^d$ Observed color at early epochs. The Galactic extinction is corrected with an extinction to reddening ratio $R_V$ of 3.1. $t$ corresponds to the phase (rest frame) relative to the $B$-band peak derived from the best-fit SALT2 model. Numbers in parentheses are 1$\sigma$ uncertainties in units of 0.01 magnitude.\\
}
\end{deluxetable*}

\begin{deluxetable*}{cccccccccc}
\movetabledown=3000mm
%\centerwidetable
\tablenum{2}
\tablecaption{Fitting Parameters of HSC-SSP SN Ia Light Curves\label{tab:Fitting Parameters}}
\tablewidth{0pt}
\tablehead{
\colhead{Name} & \colhead{$x_1$} & \colhead{$c$} & \colhead{ndf} & \colhead{$\chi^2$/ndf} & \colhead{$n_g$} & \colhead{$n_r$} & \colhead{$n_i$} & \colhead{$n_z$}
}
\startdata
HSC17bmhk & 0.29 $\pm$ 0.07 & 0.04 $\pm$ 0.02 & 16 & 2.78 & 1.40 & 1.35 & 1.37 & 1.35 \\
HSC17bqai & 0.20 $\pm$ 0.10 & 0.07 $\pm$ 0.02 & 18 & 2.40 & 1.40 & 1.32 & 1.35 & 1.30 \\
HSC17dbjm & -0.56 $\pm$ 0.10 & 0.07 $\pm$ 0.02 & 13 & 1.54 & 1.74 & 1.49 & 1.61 & 1.37 \\
HSC17bobn & -0.35 $\pm$ 0.11 & 0.16 $\pm$ 0.03 & 12 & 1.21 & 1.45 & 1.45 & 1.49 & 1.43 \\
HSC17cpaq & 0.79 $\pm$ 0.17 & 0.06 $\pm$ 0.02 & 20 & 2.10 & 1.34 & 1.27 & 1.27 & 1.25 \\
HSC17dhzs & -0.24 $\pm$ 0.35 & -0.06 $\pm$ 0.03 & 8 & 1.67 & 1.66 & 1.41 & 1.47 & 1.33 \\
HSC17dkay & -1.25 $\pm$ 0.23 & 0.11 $\pm$ 0.03 & 10 & 2.28 & 1.73 & 1.79 & 1.76 & 1.42 \\
\enddata
\tablecomments{\\
Columns 2--5 correspond to best-fit parameters ($x_1$ and $c$), the number of degrees of freedom (ndf), and the reduced $\chi^2$ for the SALT2 light-curve fittings of HSC-SSP early-phase SNe Ia. The last four columns are best-fit power-law indexes of single-band ($g$, $r$, $i$, and $z$) early-phase light curves of the HSC-SSP SNe Ia.\\
}
\end{deluxetable*}

\begin{figure*}
\gridline{\fig{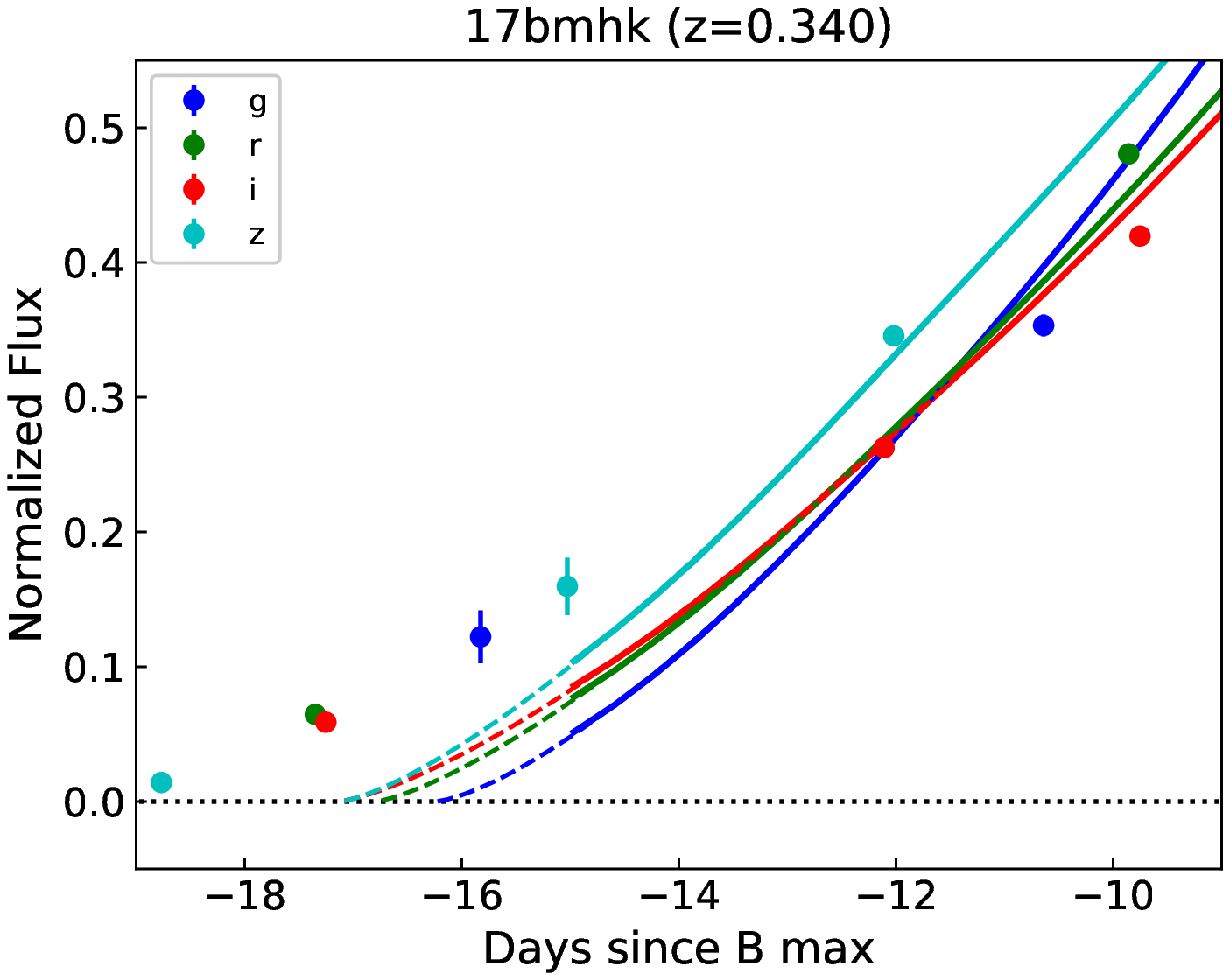}{0.6\textwidth}{(a)}
          }
\gridline{\fig{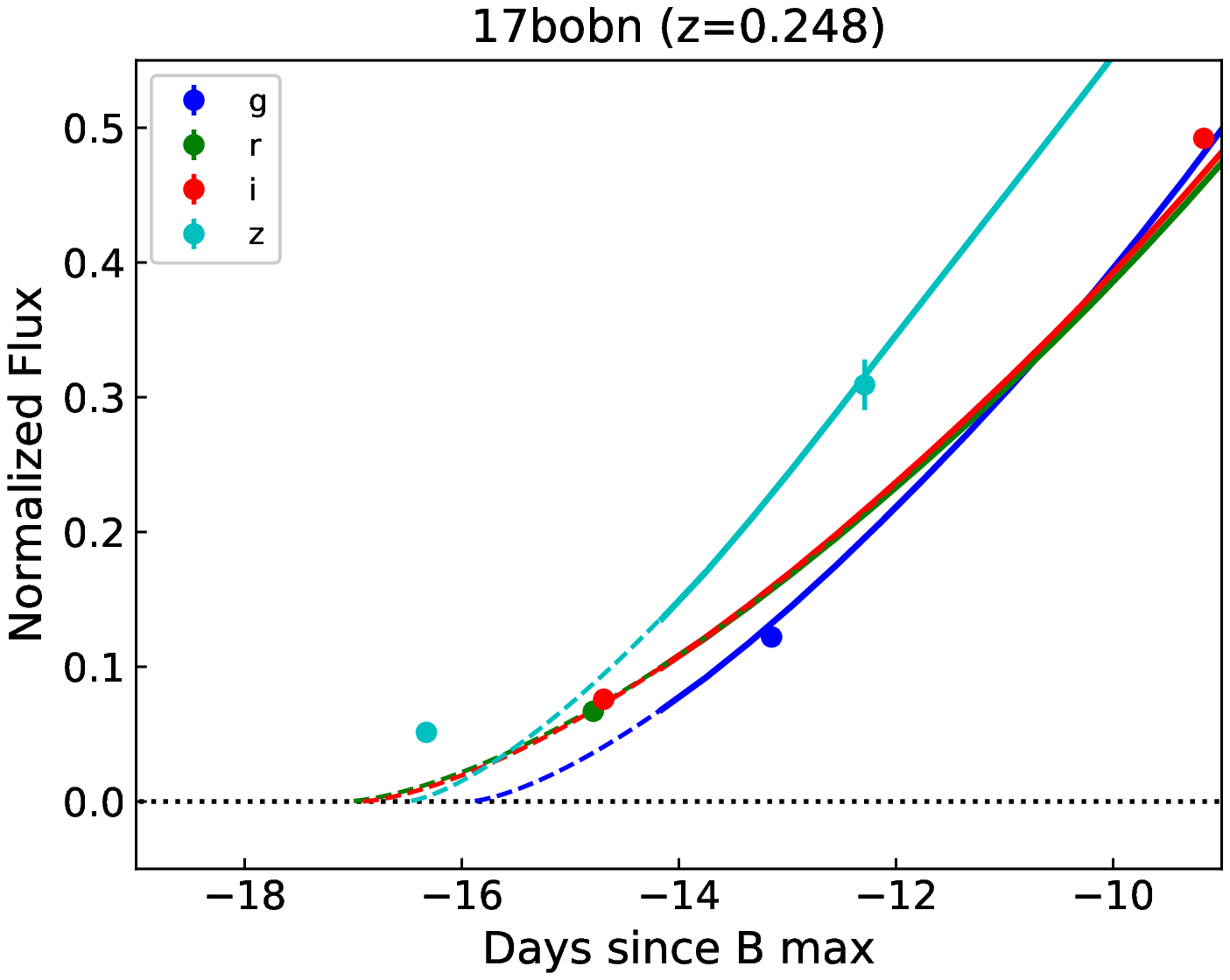}{0.34\textwidth}{(b)}
          \fig{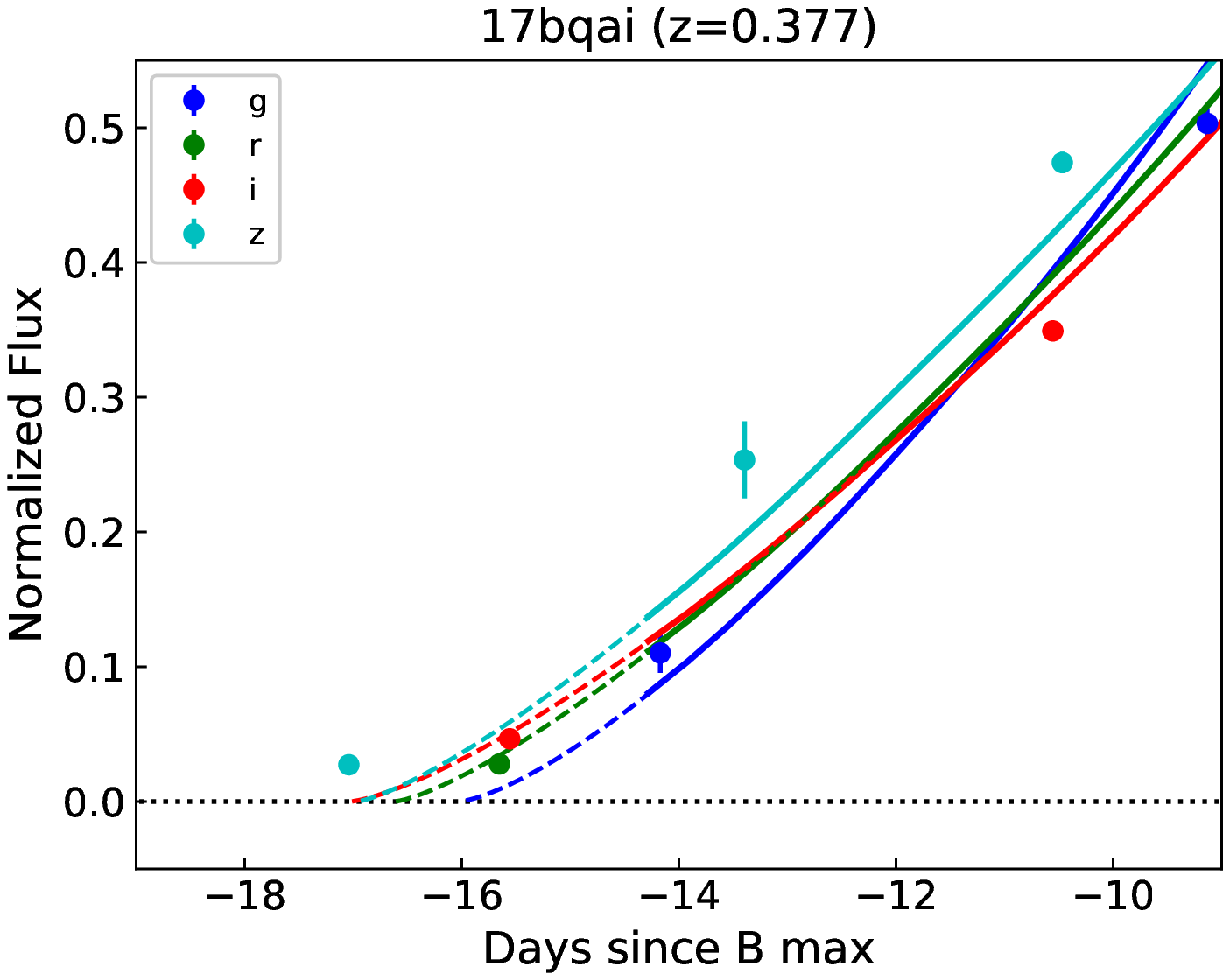}{0.34\textwidth}{(c)}
          \fig{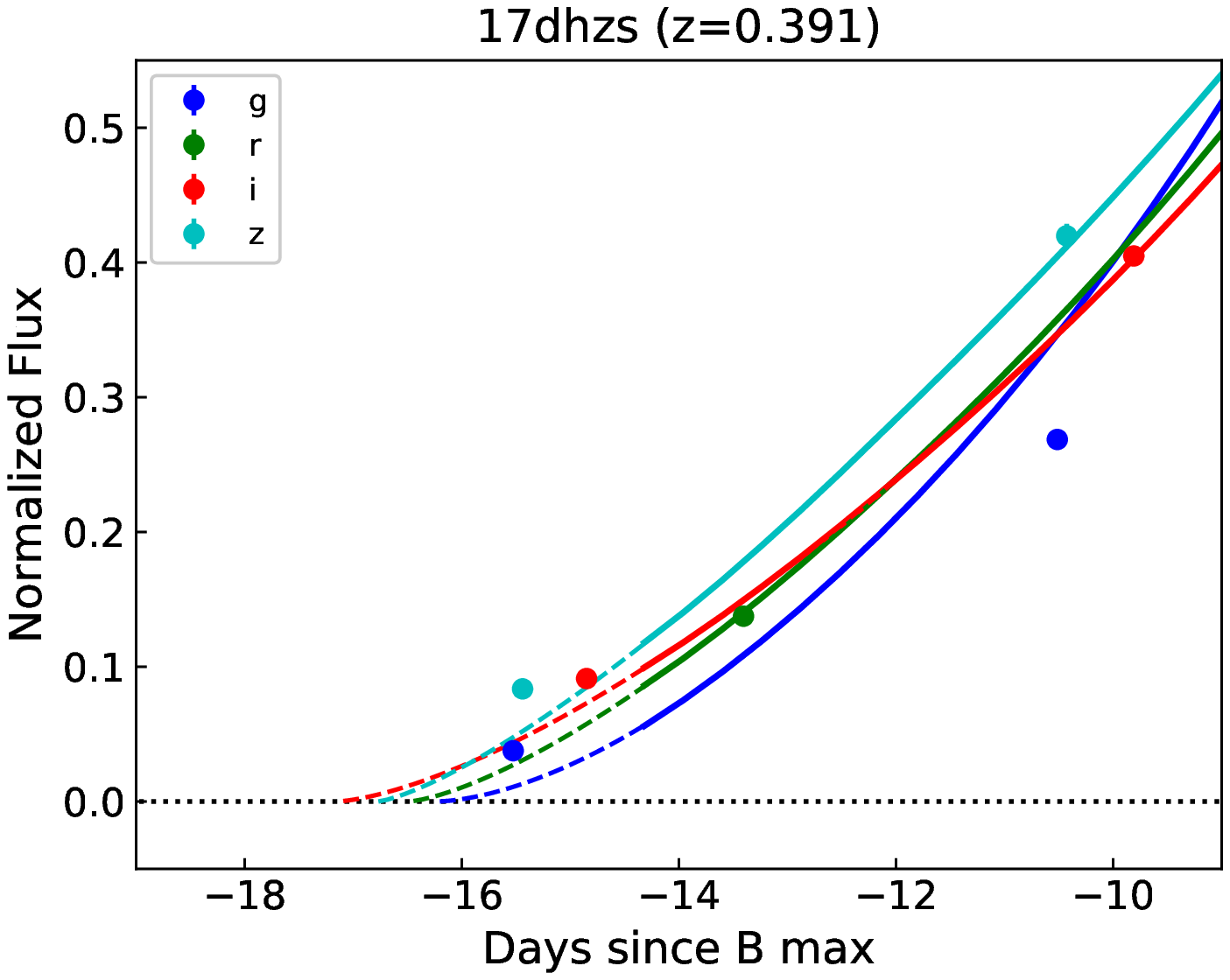}{0.34\textwidth}{(d)}
          }
\gridline{\fig{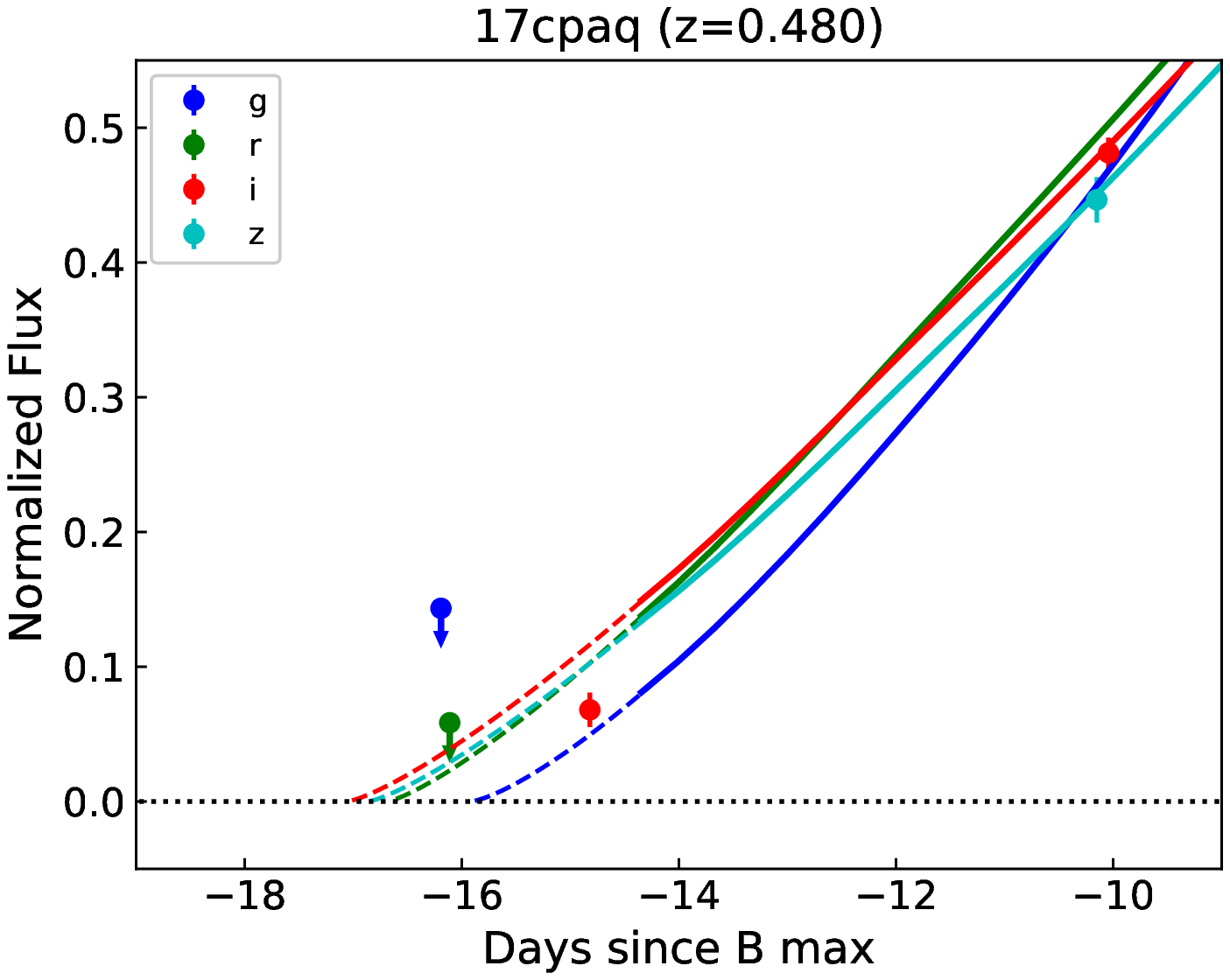}{0.34\textwidth}{(e)} 
          \fig{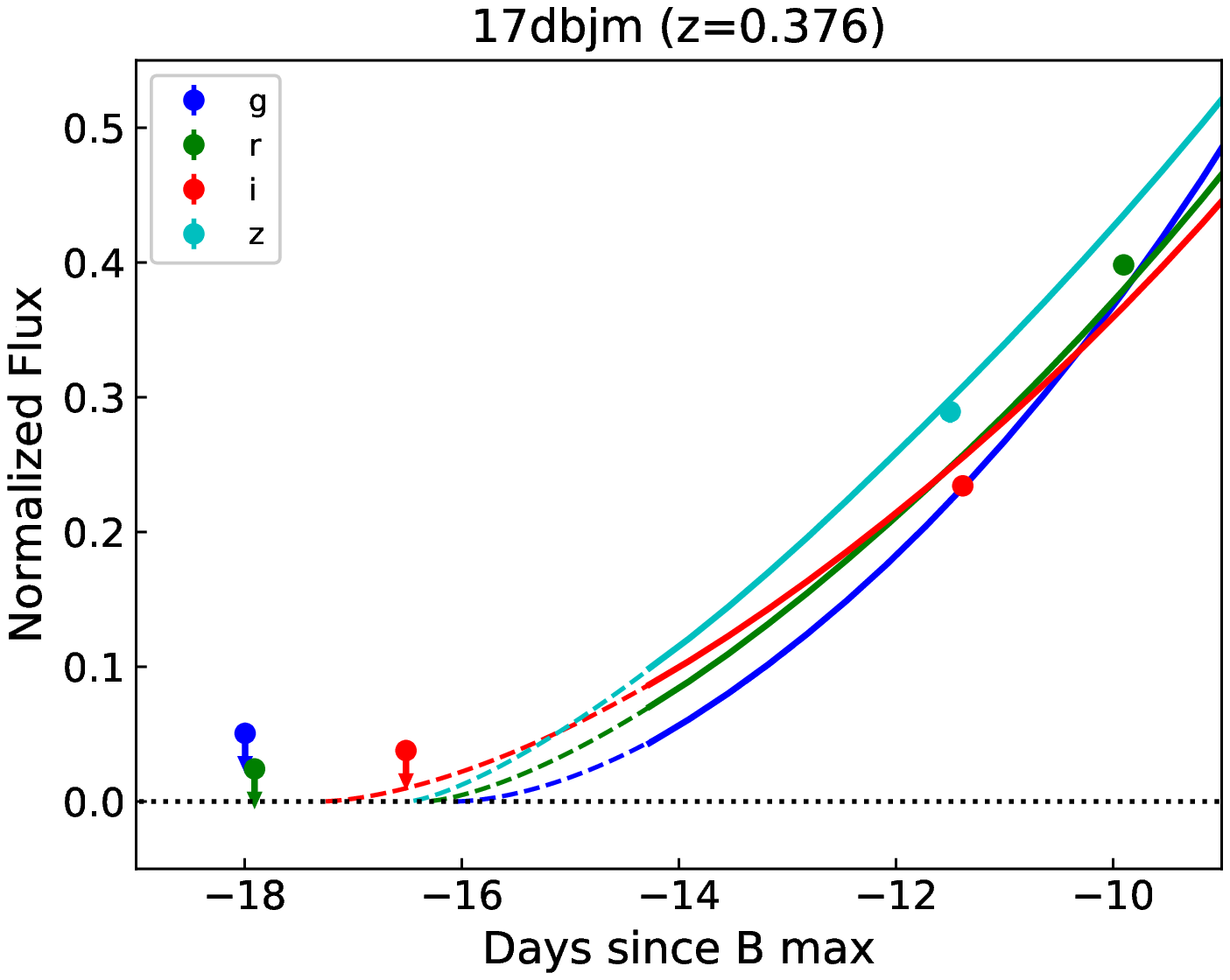}{0.34\textwidth}{(f)}
          \fig{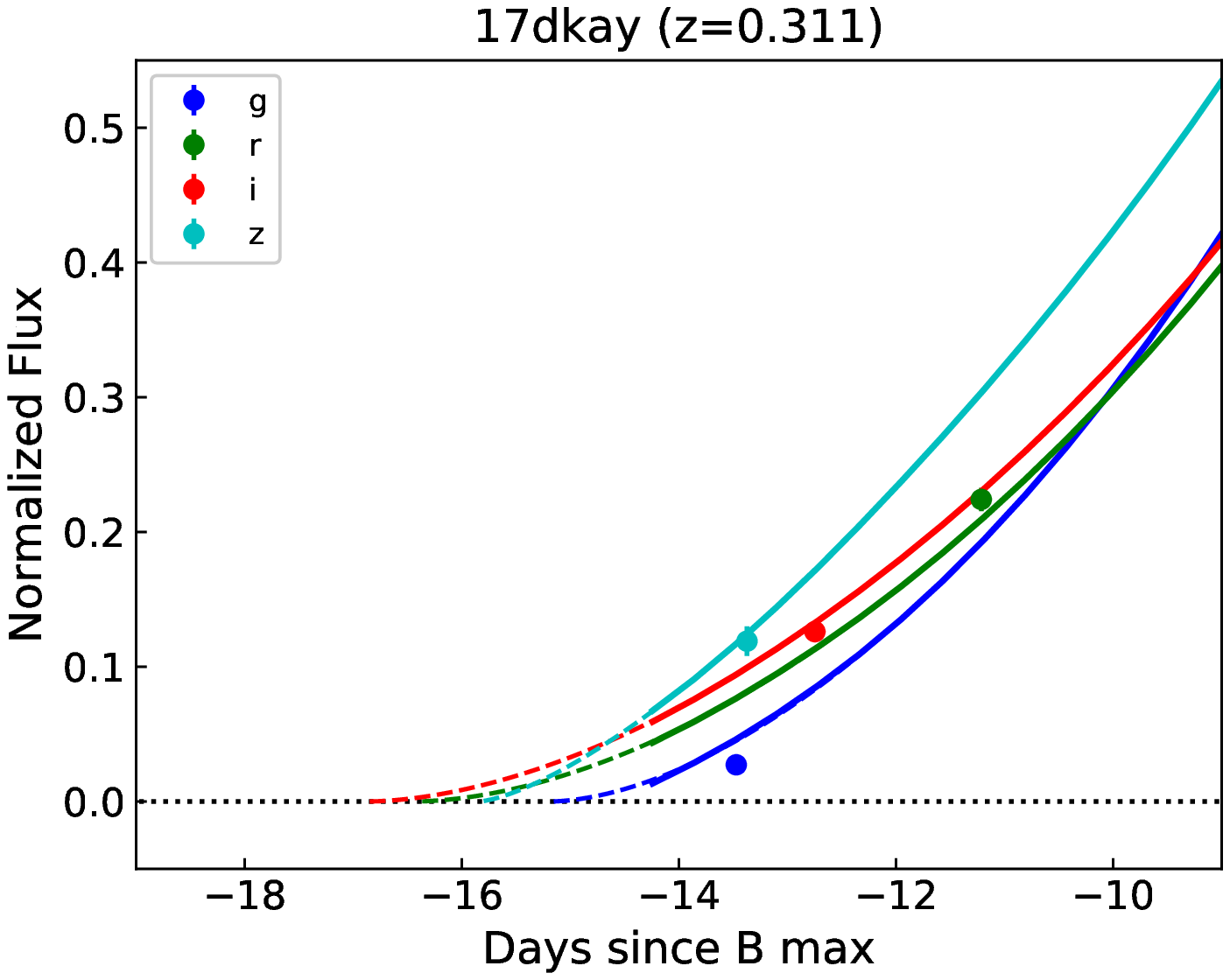}{0.34\textwidth}{(g)}
          }
\caption{First-week photometric behavior of HSC-SSP early-phase SNe Ia. Dashed curves are extrapolated power-law fittings based on SALT2 models (solid curves) with normalized flux $\lesssim$ 0.35. Points with arrows are 5$\sigma$ non-detections. Days since $B$-band maximum are in rest frame. \label{fig:EarlyLightCurves}}
\end{figure*}

The observed light curves together with the best-fit SALT2 models of the seven SNe Ia are shown in Figure 1. Due to the limited photometric/spectroscopic information and uncertainties of SNe Ia in the early phase, we did not use SALT2 to fit photometry data before $\sim -$14 days to the $B$-band peak. In order to estimate the rise time and make comparisons with ultra-early data taken by Subaru/HSC, we performed simple power-law fittings $\alpha(t-t_0)^{n}$ by adopting SALT2 best-fit models with normalized fluxes $\lesssim0.35$ in each band and extrapolated the power-law functions to the early phase (Figure 2; power-law fitting indexes are given in Table 2). As opposed to other HSC-SSP early-phase SNe Ia, a significant discrepancy between the power-law fitting and observation in all four bands is found for HSC17bmhk, indicating an abnormal brightening behavior in the early time.

\subsection{HSC17bmhk, a normal SN Ia with a blue excess in the early phase}

HSC17bmhk was discovered on MJD 57774.51, about $18.77\pm0.08$ days (rest frame) before the SALT2-fitted $B$-band peak epoch. As shown in Figure 3, there are two host candidates in terms of the visual distance to the SN. Data taken from the 3D-HST Treasury Program \citep{brammer12,skelton14,momcheva16} indicated that the spec-$z$ of the visually distant galaxy is 0.901, which gave an improper light-curve fitting of HSC17bmhk to any kinds of off-centered extremely luminous transients discovered so far (i.e. ``super-Chandrasekhar" SNe Ia, \citet{taubenberger17}; super-luminous SNe, \citet{moriya18}; fast-evolving luminous transients, \citet{tanaka16}). Given that the secondary peak feature in red wavelengths is commonly seen in normal SNe Ia and the perfect light-curve fitting to a normal SN Ia is obtained by setting a redshift of 0.340 (the spec-$z$ of the visually close galaxy), we conclude that the host of HSC17bmhk is the closer one. The best-fit light curve suggests a typical normal-type SN Ia with a rest-frame $B$-band peak magnitude of $-19.14\pm0.03$ and ${\Delta}m_{15}=1.04\pm0.01$. However, the light-curve behavior in the first four days after the discovery of HSC17bmhk is apparently different from that of other normal SNe Ia. 

\begin{figure*}
\plotone{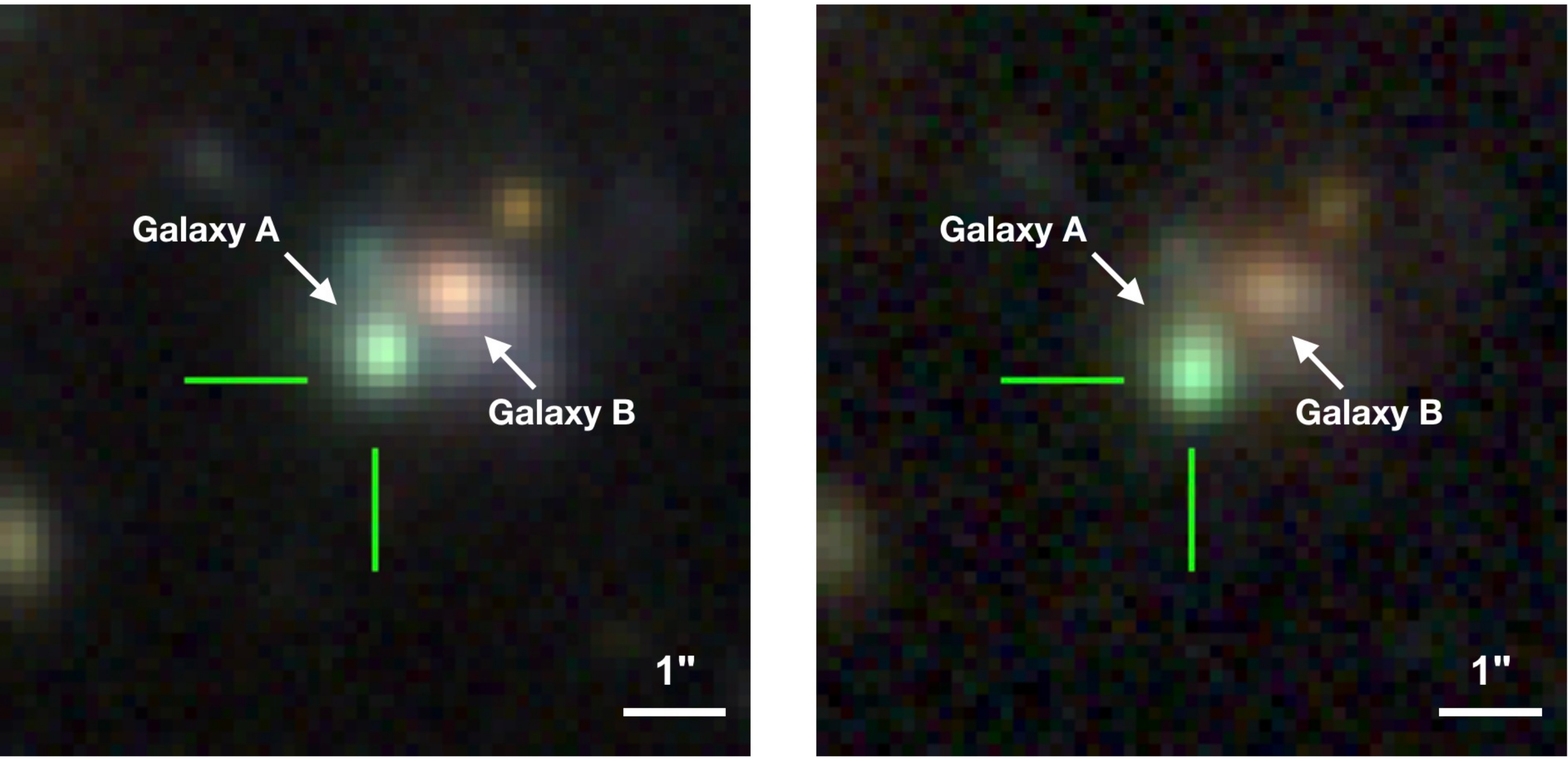}
\caption{Subaru/HSC $g$, $r$, and $i$ composite multi-color images before (left) and one week after (right) the explosion of HSC17bmhk. The image size is 10.2\arcsec $\times$ 10.2\arcsec and
the SN location is at the center. Two galaxies near HSC17bmhk are labeled in the plot. Spectroscopic redshifts of visually close (``Galaxy A") and distant (``Galaxy B") galaxies are 0.340 and 0.901, respectively. Light-curve features of HSC17bmhk indicate that the host galaxy is the closer one. \label{fig:HSC17bmhkImg}}
\end{figure*}

In Figure 2, distinct light-curve excess of HSC17bmhk can be seen in all four bands before $-15$ days to the $B$-band peak, especially in the blue wavelengths. In particular, as seen in Figure 4, the observed $g$-band normalized flux (wavelength range of which is comparable to that of the rest-frame $U$ band after taking into account the redshift effect) of $\sim0.12$ at 15.9 days before the rest-frame $B$-band peak (or 13.5 days to the observed $g$-band peak) is as high as the $U$-band excess of iPTF~14atg \citep{cao15} and slightly higher than another EExSN Ia in the normal branch, SN~2017cbv \citep{hosseinzadeh17}. The blue excess is over three times higher than the $U$-band normalized flux of the golden-normal SN Ia, SN~2011fe, at the same epoch \citep{zhang16}. In redder wavelengths, e.g. the observed $i$-band normalized flux (comparable to the rest-frame $r$ band after taking into account the redshift effect) at 17.3 days before the rest-frame $B$-band peak is similar to those of SN~2017cbv and SN~2018oh at the same epoch (Figure 4)\footnote{When we compare early photometries of different SNe Ia, it is necessary to take into account the fact that the SN Ia maximum luminosity may be correlated with its light-curve shape, and hence, the rise time. Given that previous work by \citet{firth14} indicated that the rise time before the half maximum is weakly correlated with the maximum luminosity, we chose the zero-point of time at the half peak luminosity and focus on the period before the half peak luminosity, as shown in the Figure 4.}.

\begin{figure*}
\gridline{\fig{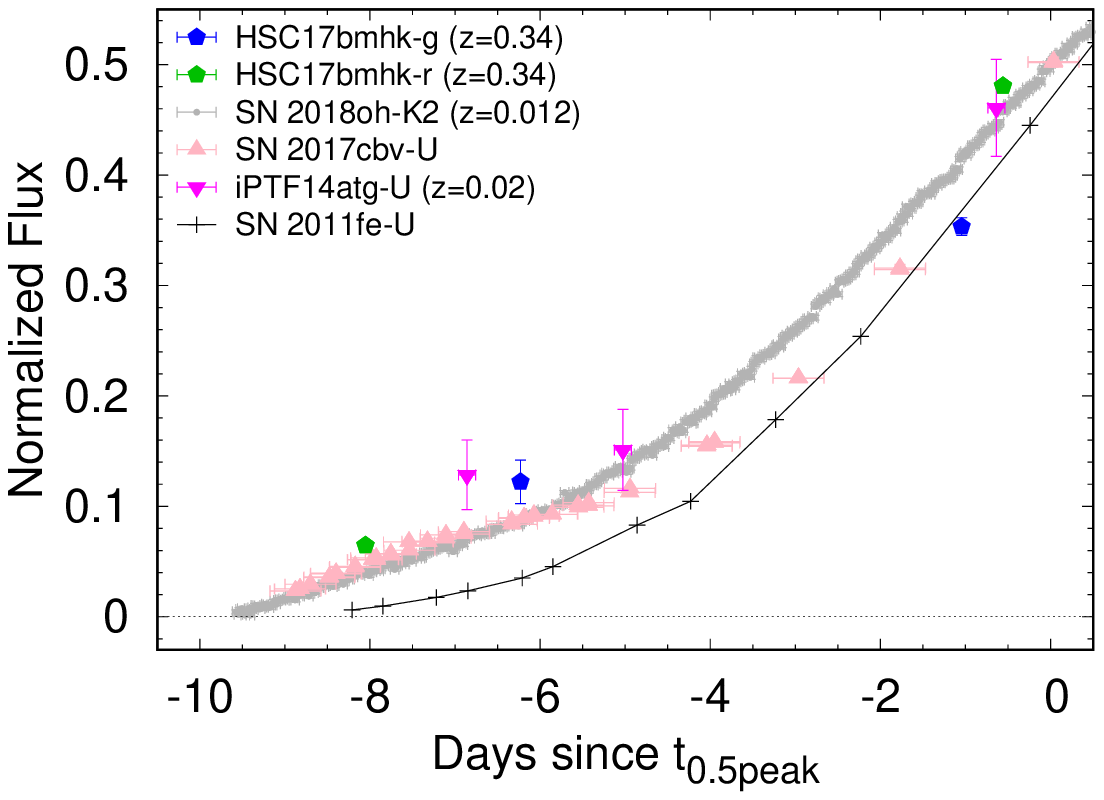}{0.5\textwidth}{(a)}
          \fig{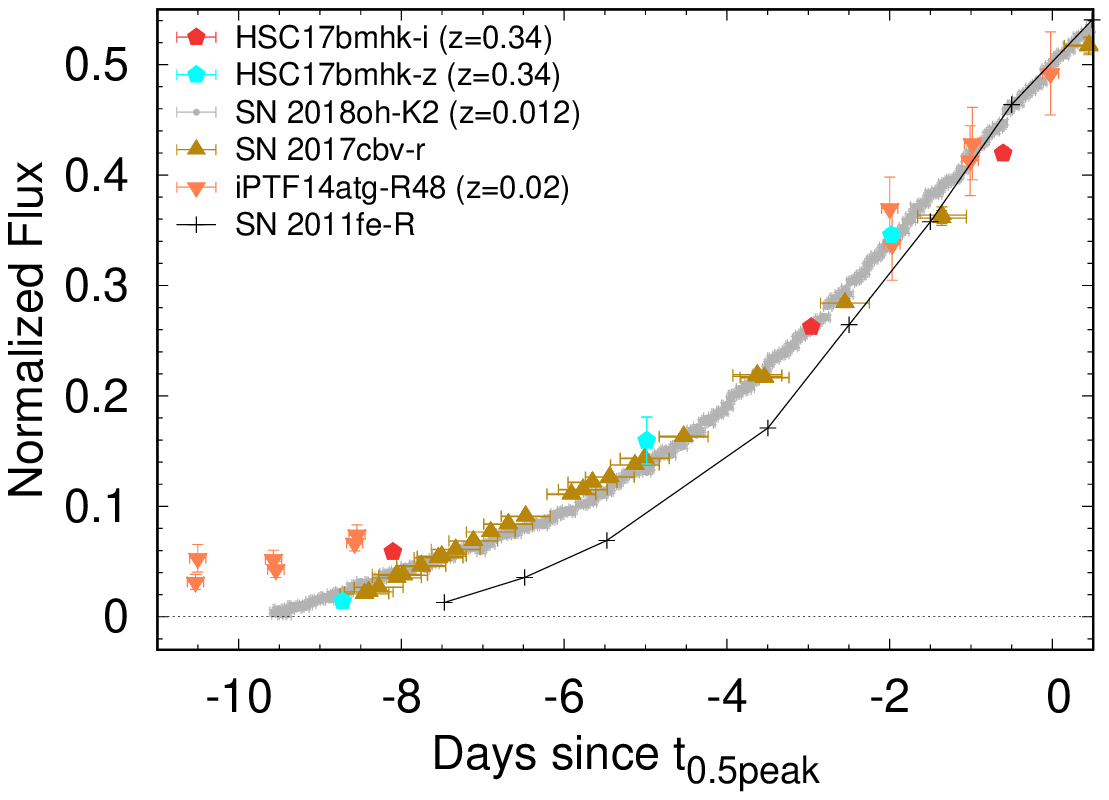}{0.5\textwidth}{(b)}
          }
\caption{Early light curves of normal SNe Ia with early excess (i.e. HSC17bmhk, SN~2017cbv, and SN~2018oh) and iPTF~14atg, a subluminous SN Ia with the most prominent early UV excess discovered so far, in blue (panel (a)) and red (panel (b)) wavelengths. A $U$-band and a $R$-band light curves of SN~2011fe (solid black lines, \citealt{zhang16}) are shown in panels (a) and (b), respectively. Days since the time at the half peak luminosity in each band (t$_{0.5peak}$) are in rest frame.\label{fig:EarlyLightCurveComparisons}}
\end{figure*}

The observed $r-i$ color\footnote{The observed color is in AB magnitude, which is approximately comparable to $g-r$ color in rest frame. The Galactic extinction is corrected, assuming an extinction to reddening ratio $R_V$ of 3.1 \citep{schlegel98}.} of $-0.14\pm0.10$ at 17.3 days before the $B$-band peak is similar to those of SN~2017cbv and Zwicky Transient Factory (ZTF) SNe Ia with blue early-color evolution (discovered in both luminous and normal branches, \citealp{bulla20}) at the same epoch. In addition, given that the absolute magnitude of SN~2017cbv is likely fainter than the one given by \citet{hosseinzadeh17} in terms of the large uncertainty of the distance and a normal ${\Delta}m_{15}(B)$ value ($\sim$ 1.06, which is much larger than the typical ${\Delta}m_{15}(B)$ of 91T/99aa-like SNe Ia, \citealp{taubenberger17}), the intrinsic luminosity of the early excess of HSC17bmhk could be similar to that of SN~2017cbv. Such a blue excess was predicted by companion/CSM-interaction as well as the surface-$^{56}$Ni-decay scenarios \citep{kasen10,piro16,maeda18}.

Particularly, the earliest photometry ($\sim4.63$ mag fainter than the peak in $z$ band) at 18.77 days before the $B$-band peak indicates a significantly longer rise time of HSC17bmhk\footnote{The rise time of HSC17bmhk is defined as the rest-frame time from HSC discovery to its $B$-band peak, which is shorter than the real value. Given that any early-excess scenarios can elongate the rise time while their influences on the peak luminosity and the light-curve shape around/post maximum are negligible, we use the non-stretch-corrected rise time to avoid possible underestimations of rise times of ``long-rising" normal SNe Ia in this paper. Stretch-corrected rise times of the HSC-SSP SNe Ia are given in Table 1 for reference.} than the typical rise time of normal SNe Ia \citep{hayden10,bianco11,ganeshalingam11,gonzalez12,JJA2018}. Such a long rise time is in favor of a hypothesis that a long dark phase may commonly exist in the normal SN Ia (\citealp{piro13,mazzali14}).

\subsection{HCS-SSP non-EExSNe Ia}

The other six early-phase SNe Ia are classified as non-EExSNe Ia in terms of generally good power-law fittings and much shorter rise times than that of HSC17bmhk. Note that imperfect fittings on early $z$-band observations of a few objects are mainly due to the lack of early observations at a rest-frame wavelength range of 6100--7700 \AA~in the SALT2 templates, thus the fittings were not well performed \citep{guy07,guy10}.

As shown in Table 1, rise times of the non-EExSNe Ia are significantly shorter than that of HSC17bmhk, which is in line with statistical studies of normal SNe Ia \citep{hayden10,bianco11,ganeshalingam11}. However, given that most of early HSC observations are in red wavelengths that are not sensitive to blue light-curve excess, it is still difficult to completely rule out companion-/CSM-interaction scenarios after taking into account differences in viewing directions and companion/CSM conditions. The tightest constraint is given by the last non-detection of HSC17dbjm, for which an $i$-band 5$\sigma$ limiting magnitude of $\sim-15.8$ at $t=-16.5$ days can rule out interactions with both red-giant and massive main-sequence companions observed from the companion direction according to 1D simulations from \citet{kasen10} and \citet{maeda18}.

Thanks to the survey strategy of the HSC-SSP transient survey, four objects have color information at $\gtrsim15$ days before the $B$-band peak (Table 1). In contrast to HSC17bmhk, non-EExSNe Ia HSC17bqai and HSC17bobn are much redder in terms of the observed $r-i$ color in the early phase, which is consistent with color evolution of normal SNe Ia without early-excess features \citep{stritzinger18,bulla20}. In addition, colors of HSC17bqai and HSC17dhzs at $t\sim-15.5$ days are too red to be produced by interacting with red-giant companions through the companion-interaction scenario \citep{kasen10,maeda18}.

As expected, the faintest SN Ia in our sample, HSC17dkay with a $B$-band peak absolute magnitude of $\sim-18.5$, shows the fastest light-curve evolution among all SNe. Even though we classify HSC17dkay as a photometrically normal SN Ia, the fast-evolving light curve and relatively low brightness compared with those of typical normal SNe Ia are also reminiscent of the so-called transitional type between normal and subluminous SNe Ia. Notably, fast-rising light curves have been found in the transitional SNe Ia \citep{yamanaka14,hsiao15}. Statistics of the rising behavior of transitional and normal SNe Ia may answer the physical relation between the two SN Ia branches.

\section{Discussion and Conclusions}

\subsection{The early excess of HSC17bmhk}

Among four early-excess scenarios (i.e. He-det, companion-interaction, CSM-interaction, and surface-$^{56}$Ni-decay), the blue and broad early excess of HSC17bmhk contradicts to the prediction of radiation from short-lived radioactive products generated by the He-det scenario \citep{JJA2017,maeda18}. As discussed in \citet{maeda18}, although both companion-interaction and CSM-interaction scenarios can generate such a blue excess by tuning the progenitor system configuration and CSM distribution, respectively, the CSM-induced early excess evolves generally faster due to a short characteristic timescale of the evolution in the bolometric light curve of CSM-induced early excess. Note that even though the timescale of CSM-induced early excess can become comparable to that of HSC17bmhk by adopting a CSM mass of $0.6~M_{\odot}$, as shown in panel (a) of Figure 5, the rate of merging WD--WD progenitor systems with a total mass of $\gtrsim 2~M_{\odot}$ is below 5\% of the Galactic SN Ia rate according to population synthesis \citep{ablimit16}. In addition, such a system may suffer from a violent merger ignition \citep{tanikawa15,sato15} where massive CSM are hardly formed in time. An even better fitting on the early excess of HSC17bmhk can be achieved by interacting with a typical main-sequence companion at $\sim2\times10^{12}$ cm away from the explosion site through the companion-interaction channel (panel (b) in Figure 5). However, given that the fitting is based on nonconsecutive photometries without very early HSC $g$-band information and a large uncertainty between 1D and 2D companion-interaction simulations \citep{kasen10,kutsuna15}, we caution that the constraint from the companion-interaction fitting is limited. Moreover, the early blue bump was also reproduced through the surface-$^{56}$Ni-decay channel and the early excess of SN~2018oh, another HSC17bmhk-like EExSN Ia, was fitted by assuming a near-Chandrasekhar mass WD explosion with a $0.03~M_{\odot}$~$^{56}$Ni surface layer, though its early-color evolution cannot be perfectly reproduced by any early-excess scenario \citep{dimitriadis19}. Therefore, other independent indicators are necessary to figure out the early-excess origin of HSC17bmhk.

\begin{figure*}
\gridline{\fig{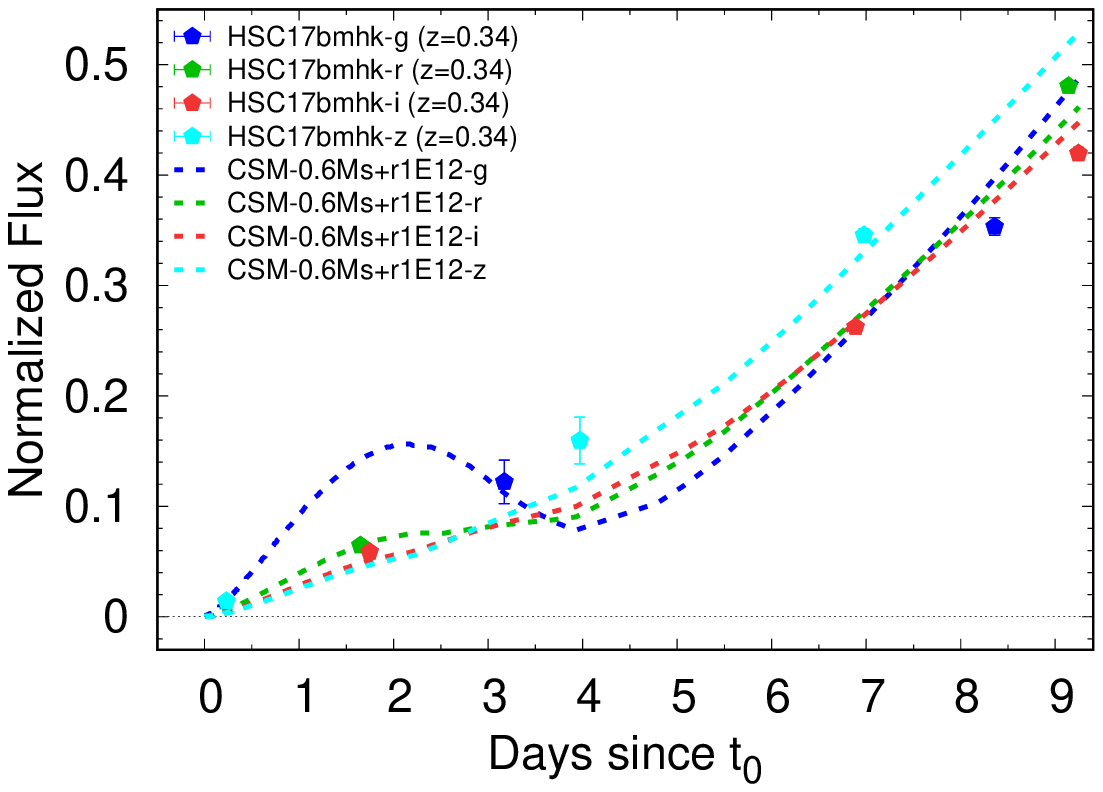}{0.5\textwidth}{(a)}
          \fig{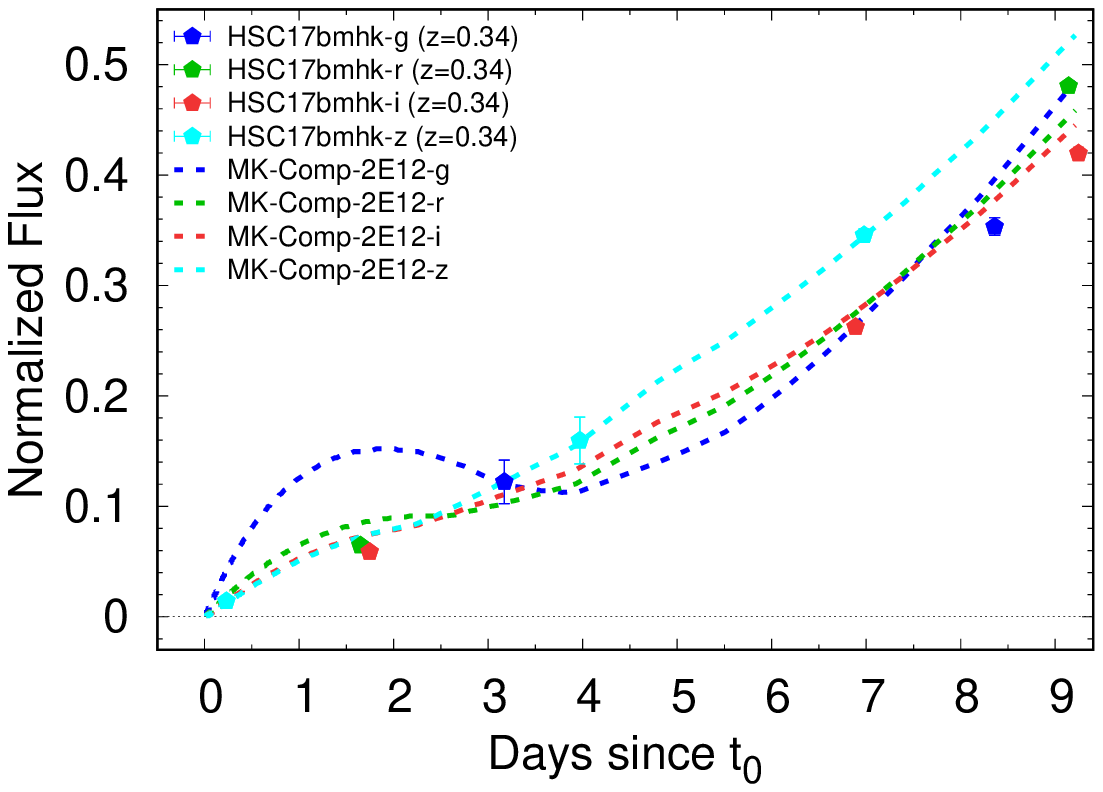}{0.5\textwidth}{(b)}
          }
\caption{Model fittings of early light curves of HSC17bmhk. The left panel shows synthesized light curves of a 1D CSM-interaction simulation assuming CSM mass of $0.6~M_{\odot}$ with a characteristic (outer edge) radius of $1.0\times10^{12}$ cm and the SALT2 best-fit light curves. The right panel shows synthesized light curves of a 1D companion-interaction simulation by assuming a non-degenerate companion at a separation distance of $2.0\times10^{12}$ cm \citep{maeda18} and the SALT2 best-fit light curves. Days since the explosion time ($t_0$) are in rest frame. \label{fig:EarlyLightCurveFittings}}
\end{figure*}

\subsection{Implications from the rise time of normal SNe Ia}

The rise time of $>18.8$ days of HSC17bmhk is significantly longer than the typical rise time of normal SNe Ia \citep{hayden10,bianco11,gonzalez12}. Although a scatter of rise times of normal SNe Ia has been noticed by previous studies \citep{ganeshalingam11,firth14}, there is no systematical investigation of normal SNe Ia with long rise times due to the rarity and poor data quality of such objects. Recently, thanks to a booming number of early-phase SNe Ia discovered by transient surveys with day or even shorter cadences, it turns out that the fraction of such ``long-rising" normal SNe Ia is non-negligible. For instance, by investigating normal SNe Ia with $g$-band rise times more than 18 days (rest frame, all SNe Ia with observed $g$-band absolute magnitudes between $-19.0$ and $-19.4$) discovered by ZTF \citep{yao19}, over half (ZTF18abkhcwl, ZTF18aavrwhu, ZTF18abxxssh, ZTF18abpamut, ZTF18aaqqoqs, and ZTF18abimsyv) show possible early excess in their $g$-band light curves, indicating an extended rise-time distribution of normal SNe Ia and a high early-excess fraction of the ``long-rising" normal SNe Ia. The long rise times discovered in a fraction of normal SNe Ia suggests that many normal SNe Ia may have dark phases lasting for few days after explosions, and the timescale from the SN explosion to the peak can be very similar among normal SNe Ia.

Physically, the observed rise-time dispersion of normal SNe Ia can be explained by a scatter of the $^{56}$Ni distribution in the SN ejecta \citep{piro13,piro16}. The ``long-rising" normal SN Ia requires a more $^{56}$Ni-abundant surface, making it reminiscent of luminous SNe Ia that commonly show blue early excess and longer rise times likely due to the massive $^{56}$Ni distribution at the outermost layer of SN ejecta \citep{JJA2018}. Observationally, similar early light-curve behavior and comparable rise times are found for EExSNe Ia in both luminous and normal branches, suggesting the same physical origin of the blue-bump features discovered in the two SN Ia subclasses. Moreover, relatively shallow Si \textsc{ii} adsorptions found in both SN~2017cbv and SN~2018oh in the early phase may also imply efficient detonations at surfaces of their progenitors.

Current light-curve simulations of the surface-$^{56}$Ni-decay scenario simply investigate the light-curve behavior by tuning the $^{56}$Ni fraction at different layers manually. Therefore, the physical reality of such toy models is still an open question. Phenomenologically, in order to generate the early-excess feature, an inhomogeneous $^{56}$Ni distribution may be required \citep{magee20}. Figuring out the physical mechanism of producing such an inhomogeneous $^{56}$Ni distribution is beyond the scope of this paper. As a possible solution, we infer that an asymmetrical explosion of the progenitor \citep{maeda11,livneh19} may give rise to the uneven $^{56}$Ni distribution reaching to the surface of the ejecta, and the rise-time scatter can be naturally explained by taking into account the viewing-angle effect. A mean power-law fitting index of $\sim$ 2 may also be expected, which was discovered by a very recent work with a more unbiased early SN Ia sample from ZTF \citep{miller20}. Future multi-denominational hydrodynamic simulations will answer whether or not the large scatter of early-phase light curves can be explained by the off-center explosions.

A day-cadence survey in red wavelengths (or without filters) may still have a difficulty to distinguish such inconspicuous blue bumps. Note that two previously reported HSC17bmhk-like EExSNe Ia, SN~2017cbv and SN~2018oh, were discovered with hours/minutes-cadence observations. We infer that normal SNe Ia with long rise times may commonly show the surface-$^{56}$Ni-decay-induced blue excess as found in HSC17bmhk and other EExSNe Ia in the normal branch, which can be easily identified with early UV photometries or high-cadence (hours or shorter) blue-band surveys.

\subsection{Summary}

In this paper, we reported seven early-phase photometrically normal SNe Ia discovered by the HSC-SSP transient survey in 2017, including one EExSN Ia, HSC17bmhk, which shows a blue light-curve excess in the first four days after the discovery. Although the early excess of HSC17bmhk can be explained by both interactions with a non-degenerate companion or surrounding dense CSM and radiation powered by radioactive decays of abundant $^{56}$Ni at the surface of the SN ejecta, theoretical predictions based on sparse early-phase photometries cannot provide a stringent constraint on the origin of the early excess of HSC17bmhk. In order to identify the origin of such a blue excess and thus figure out the evolutionary pathway leading to normal SNe Ia, it is important to systematically study a sample of SNe Ia in terms of the rise time and the blue early excess.

A growing number of normal SNe Ia with long rise times have been discovered, indicating that a part of normal SNe Ia have $^{56}$Ni-abundant outer layers of their ejecta. Given a similarity in the nature of the blue excess to a luminous SN Ia subclass and a possible high early-excess fraction for the ``long-rising" normal SNe Ia, the surface-$^{56}$Ni-decay scenario is the most promising early-excess scenario to explain the blue early excess in the two SN Ia branches. To achieve the scatter of the $^{56}$Ni distribution (i.e. the scatter of the rise time) in theory, a thermonuclear explosion triggered at an offset from the center of the WD progenitor is likely required. Statistically, with a large number of early-phase SNe Ia discovered by ongoing and upcoming high-cadence blue-band transient surveys, the fraction of normal SNe Ia with early excess similar to that of HSC17bmhk is expected to be higher than that predicted by the companion-interaction scenario.

\vspace{10pt}

We thank the anonymous referee for helpful comments and suggestions. This work has been supported by the Japan Society for the Promotion of Science (JSPS) KAKENHI grant 19K23456 and 18J12714 (J.J.), 18H04342 and 16H01087 (J.J. and M.D.), 18H05223 (J.J., K.M., M.D., and T.S.), 18H04585 and 17H02864 (K.M.), 15H02075, 16H02183, and 19H00694 (M.T.), 18K03696 (N.S.), 16H06341, 16K05287, and 15H02082 (T.S.), and MEXT 17H06363 (M.T.). I.T., N.S., and N.Y. acknowledge financial support from JST CREST (JPMHCR1414).

The Hyper Suprime-Cam (HSC) collaboration includes the astronomical communities of Japan and Taiwan, and Princeton University. The HSC instrumentation and software were developed by the National Astronomical Observatory of Japan (NAOJ), the Kavli Institute for the Physics and Mathematics of the Universe (Kavli IPMU), the University of Tokyo, the High Energy Accelerator Research Organization (KEK), the Academia Sinica Institute for Astronomy and Astrophysics in Taiwan (ASIAA), and Princeton University. Funding was contributed by the FIRST program from Japanese Cabinet Office, the Ministry of Education, Culture, Sports, Science and Technology (MEXT), the Japan Society for the Promotion of Science (JSPS), Japan Science and Technology Agency (JST), the Toray Science Foundation, NAOJ, Kavli IPMU, KEK, ASIAA, and Princeton University. 

This paper makes use of software developed for the Large Synoptic Survey Telescope. We thank the LSST Project for making their code available as free software at http://dm.lsst.org

The Pan-STARRS1 Surveys (PS1) have been made possible through contributions of the Institute for Astronomy, the University of Hawaii, the Pan-STARRS Project Office, the Max-Planck Society and its participating institutes, the Max Planck Institute for Astronomy, Heidelberg and the Max Planck Institute for Extraterrestrial Physics, Garching, The Johns Hopkins University, Durham University, the University of Edinburgh, Queen’s University Belfast, the Harvard-Smithsonian Center for Astrophysics, the Las Cumbres Observatory Global Telescope Network Incorporated, the National Central University of Taiwan, the Space Telescope Science Institute, the National Aeronautics and Space Administration under Grant No. NNX08AR22G issued through the Planetary Science Division of the NASA Science Mission Directorate, the National Science Foundation under Grant No. AST-1238877, the University of Maryland, and Eotvos Lorand University (ELTE) and the Los Alamos National Laboratory.

This research is based in part on data collected at the Subaru Telescope and retrieved from the HSC data archive system, which is operated by the Subaru Telescope and Astronomy Data Center at NAOJ.

This work is based on observations taken by the 3D-HST Treasury Program (GO 12177 and 12328) with the NASA/ESA HST, which is operated by the Association of Universities for Research in Astronomy, Inc., under NASA contract NAS5-26555.

\vspace{10pt}

\facility{Subaru (HSC)}

\software{hscPipe \citep{bosch18}, LSST pipeline \citep{axelrod10,ivezic19}, SALT2 \citep{guy07,guy10}, Astropy \citep{astropy13,astropy18}}

\bibliographystyle{aasjournal}
%\bibliography{Bibliography_HSC-SSP_ESNeIa_Jiang.bib}
\bibliography{main.bbl}

\end{document}